\documentclass[a4paper,12pt]{article}
\usepackage[round]{natbib}
\usepackage{amssymb,lscape,psfrag}
\usepackage{epsfig,color}
\usepackage{array}
\usepackage[round]{natbib}
\usepackage{enumerate}
\usepackage{amsmath}
\usepackage{graphics}
\usepackage{graphicx}
\usepackage{epsfig}
\usepackage{verbatim}
\usepackage{cancel}
\usepackage{longtable}
\usepackage{multirow}
\usepackage{float}

\begin{document}
\newcommand {\inispace}{\renewcommand{\baselinestretch}{1.}\normalsize}
\newcommand {\tabspace}{\renewcommand{\baselinestretch}{1.}\normalsize}

\inispace

\title{Bayesian inference for transportation origin-destination matrices: the Poisson-inverse Gaussian and other Poisson mixtures}
\date{}
\author{Konstantinos Perrakis\footnote{Transportation Research Institute,
   Hasselt University, Belgium (1st author now in: Department of Hygiene, Epidemiology and Medical Statistics, Medical School, University of Athens, Greece); kperrakis@med.uoa.gr, davy.janssens@uhasselt.be},
  Dimitris Karlis\footnote{Department of Statistics, Athens University of
    Economics and Business, Greece; karlis@aueb.gr},
Mario Cools\footnote{LEMA, University of Li\`{e}ge, Belgium; mario.cools@ulg.ac.be}\\ and
Davy Janssens$^*$}

\maketitle

\begin{abstract}

\noindent In this paper we present Poisson mixture
approaches for origin-destination (OD) modeling in transportation
analysis. We introduce covariate-based models which incorporate
different transport modeling phases and also allow for direct
probabilistic inference on link traffic based on Bayesian
predictions. Emphasis is placed on the Poisson-inverse Gaussian as
an alternative to the commonly-used Poisson-gamma and
Poisson-lognormal models. We present a first full Bayesian
formulation and demonstrate that the Poisson-inverse Gaussian is
particularly suited for OD analysis due to desirable marginal and
hierarchical properties. In addition, the integrated
nested Laplace approximation (INLA) is considered as an alternative to Markov chain Monte Carlo and the two methodologies are compared under specific modeling assumptions. The case study is based on 2001 Belgian
census data and focuses on a large, sparsely-distributed OD matrix
containing trip information for 308 Flemish municipalities.

\vspace{0.3cm} \noindent \textit{Keywords:} Hierarchical Bayesian
modeling; INLA; OD matrix; overdispersion; Poisson mixtures

\end{abstract}

\section{Introduction}

In transportation analysis the \textit{travel demand} within a
geographical area, dividable into a given number of
non-overlapping zones, is summarized by an OD matrix which
contains the \textit{trips} or \textit{flows }that have
occurred from each zone of that area to every other zone. Consider
an area which can be divided into $m$ zones and let $T_{od} $
denote the flows from zone of \textit{origin }$o$ to
zone of \textit{destination }$d$, where $o,d = 1,2,...,m$. The OD
matrix ${\rm {\bf T}}$, is then
\[
{\rm {\bf T}} = \left[ {\begin{array}{cccc}
 T_{11} &T_{12}& \ldots &T_{1m} \\
 T_{21} &T_{22}&\ldots & T_{2m} \\
 \vdots & \vdots & \ddots& \vdots \\
 T_{m1} &T_{m2} &\ldots&T_{mm} \\
 \end{array}} \right].
\]
The elements $T_{od}$, for $o \ne d$,  correspond to
\textit{inter-zonal} flows, whereas the elements across the main
diagonal $T_{oo}$ correspond to \textit{intra-zonal} flows. The marginal totals
$T_{o \bullet}=\sum_{d}T_{od}$ and $T_{\bullet d}=\sum_{o}T_{od}$ are commonly
referred to as \textit{trip-productions} and \textit{trip-attractions}, respectively.
In lexicographical order the matrix ${\rm {\bf T}}$ can be represented as
${\rm {\bf y}} = (y_1 ,y_2,\ldots,y_n )^T \\ \equiv (T_{11} ,T_{12},\ldots,T_{mm} )^T$ with $n = m^2$.

The inferential scope in OD modeling depends on several defining
aspects such as spatial resolution, time resolution
and classification by
trip-purpose.
In
addition,
OD modeling is itself part of a larger inferential
framework.
Specifically,
the traditional transportation modeling framework consists of a
sequence of 4 modeling steps, namely (a) trip-generation, (b)
trip-distribution, (c) modal-split and (d) traffic-assignment.
Trip-generation models are typically regression or cross-classification models which relate trip-productions and trip-attractions to socio-economic, location and land-use variables.
Trip-distribution models balance trip-productions and trip-attractions, and distribute the trips to the cells of an OD matrix
usually by using supplementary prior information in the form of an outdated OD matrix. Commonly used trip-distribution models include gravity and direct-demand models.
The subsequent step of modal-split entails disaggregating the OD matrix with
respect to mode choice. Finally, traffic-assignment
involves
allocating the $n - m$ inter-zonal flows on a corresponding
transport network consisting of all the available links
which define the possible routes from zone of origin $o$
to zone of destination $d$, for $o,d = 1,2,...,m$ and $o \ne d$.
Interested readers are
referred to Ort\'{u}zar and Willumsen (2001) for four-step modeling and to Thomas (1991) for traffic-assignment.

In general, the four-step
procedure remains widely accepted by transportation planners, so that OD modeling up to the present is mainly based on
trip-generation and trip-distribution principles.
A first modern Bayesian approach to trip-distribution,
based on the gravity model, is discussed in West (1994). It is also worth noting that a
different approach for OD estimation relies on information from
link traffic data where the traffic-assingment problem is actually
inverted; see e.g. Tebaldi and West (1998) and Hazelton (2010) for
Bayesian methods. The methodological framework is quite different under this approach
and it is actually part of a broader literature on
network tomography (e.g. Medina et al., 2002).
In this study we extend the methodology presented in Perrakis et al. (2012a) for OD modeling based on census data and Perrakis et al. (2012b) for traffic-assignment inference through Bayesian predictions. Additional references concerning OD estimation from travel-surveys and/or link traffic can be found in Perrakis et al. (2012a).

In particular, we investigate the performance of three Poisson
mixture models, namely the Poisson-gamma (PG), Poisson-lognormal
(PLN) and Poisson-inverse Gaussian (PIG) models. The PG model is
the most commonly used and well established model within the
family of Poisson mixtures, while the PLN model remains up to
present the predominant alternative. The PIG model is the less
known and less used model among the three, especially within the
Bayesian framework. We present a first full Bayesian treatment of
the PIG model and demonstrate that it has desirable properties
both in its marginal and in its hierarchical forms. In addition,
we consider the integrated nested Laplace approximation (INLA)
framework (Rue et al., 2009) as a potentially efficient
alternative to Markov chain Monte Carlo methods for the PG and PLN
models. The case study focuses on a large-scale OD matrix, derived
from the 2001 Belgian census study, containing trip-information
for 308 municipalities in the region of Flanders.

The paper is organized as follows.  Literature review and Bayesian formulations for the three models in question are provided in section 2. The OD matrix, the transport
network of Flanders and the selection of explanatory variables are
described in section 3. Results are
presented in section 4. The paper ends with conclusions and
considerations of future research in section 5.

\section{Poisson mixture models}

With Poisson mixture models we assume that the OD flows $y_i $ are
i.i.d. Poisson realizations and that the rate of the Poisson
distribution is $\lambda _i = \mu _i u_i $ for $i = 1,2,...,n$.
The rate $\lambda _i $ is split in two parts; $\mu _i $ is the
part which is related to the vector of $p + 1$ unknown parameters
${\boldsymbol {\beta }} = (\beta _0 ,\beta _1 ,...,\beta _p )^T$ and
the set of explanatory variables ${\rm {\bf x}}_i = (1
,x_{i1} ,...,x_{ip} )^T$ through the log-link function $\log \mu
_i = {\boldsymbol {\beta }}^T{\rm {\bf x}}_i $, and $u_i $ is a random
component -- interpreted as a multiplicative random effect
accounting for heterogeneity -- which is attributed with a density
$g_1 (u_i )$. The Poisson mixture modeling formulation is
summarized as follows
\[
\begin{array}{l}
 y_i \sim Pois(\lambda _i ),\mbox{ with }\lambda _i = \mu _i u_i \mbox{ and}
\\
 \mu _i = \exp({{\boldsymbol {\beta }}^T{\rm {\bf x}}_i }), \\
 u_i \sim g_1 \mbox{(}u_i \mbox{) and }E(u_i ) = 1. \\
 \end{array}
\]
The density $g_1 $ is known as the mixing density and can be
continuous, discrete or even a finite support distribution. The
constrain on the expected value of the random component $u_i $
ensures that the model is scale-identifiable. Poisson mixtures are
employed as overdispersed alternatives to the simple Poisson model
which arises when the mixing density becomes degenerate.
Alternatively, from a GLMM perspective the
above model can be expressed as
\[
\begin{array}{l}
 y_i \sim Pois(\lambda _i )\mbox{ with log}\lambda _i = {\boldsymbol {\bf \beta
}}^T{\rm {\bf x}}_i + \varepsilon _i , \\
 \varepsilon _i \sim g_2 (\varepsilon _i )\mbox{ and }E(\varepsilon _i ) =
0, \\
 \end{array}
\]
where $\varepsilon _i $ is an additive random error term. Here the
constraint on the expected value ensures location-identifiability.
The two formulations are equivalent, however the intercepts and
the interpretations of marginal means are different due to the
identifiability constraints (Lee and Nelder, 2004). The Poisson
likelihood is the conditional likelihood given the unobserved
random effect vector ${\rm {\bf u}} = (u_1 ,u_2 ,...,u_n )^T$.
Integration over ${\rm {\bf u}}$ results to the marginal sampling
likelihood, i.e. $p(\rm {\bf y \vert \boldsymbol \mu}) = \int {p({\rm {\bf y}}\vert
{\boldsymbol {\mu }},{\rm {\bf u}})g_1 ({\rm {\bf u}})d{\rm {\bf u}}}
$. Frequentist inference usually focuses on the marginal structure
under maximum-likelihood (ML),
restricted-ML, quasi-likelihood and
pseudo-likelihood estimation procedures.

When the mixing density $g_1 $ is a gamma distribution, we have
the PG model which is the most frequently used Poisson mixture
model due to the property that the resulting marginal likelihood
is a negative binomial distribution. Properties and estimation
procedures for negative binomial regression can be found in
Lawless (1987).
The PG model is also included in the family of
hierarchical generalized linear models (HGLM's) introduced by Lee
and Nelder (1996) who provide ML estimates for regression
parameters as well as random effects based on the hierarchical
likelihood (h-likelihood). The PLN model arises when $g_1 $ is a
lognormal distribution. The resulting marginal distribution of
this model, known simply as Poisson-lognormal (Shaban, 1988), does
not have a closed form expression and thus numerical integration
is needed for marginal estimation. Nevertheless, the PLN model is
regularly used in practice due to its distinct historical
development as a GLMM for count data based on the assumption that
$g_2 $ is a normal distribution
(Breslow 1984).
Estimation of the model
through Gaussian quadrature and the EM algorithm is handled in
Aitkin (1996). An inverse Gaussian (IG) density for $g_1 $ results
in the PIG model which leads to a Poisson-inverse Gaussian
marginal density. This distribution, unlike the Poisson-lognormal
case, does have a closed form expression.
The PIG model
was first presented by Holla (1967).
Information on ML estimation can be found in Dean et al. (1989) and in the references therein. The PIG model has been used in the actuarial science (Willmot, 1987; Carlson, 2002) and in linguistics where the zero-truncated marginal form is of particular interest (e.g. Puig et al., 2009).

A first consideration of all three models is presented in Chen and
Ahn (1996). Later, Karlis (2001)
provided a generally applicable EM algorithm for Poisson mixtures
and compared the three models on a real dataset. In Boucher and
Denuit (2006), the performance of the three models is investigated from
a random-effects versus fixed-effects perspective on motor
insurance claims. Finally, in Nikoloulopoulos and Karlis (2008)
the models are compared with respect to distributional properties
such as skewness and kurtosis under simulation experiments. This
study illustrates some theoretical expectations, namely that the PLN
and PIG models allow for longer right tails
and are thus more appropriate than the PG model for modeling
highly positive-skewed data.

From a Bayesian perspective, Poisson mixtures have a natural
interpretation as hierarchical or multilevel models where the
mixing distribution is considered as a first-level prior
of which the parameters are assigned
with a second-level prior or hyperprior.
With respect to the equivalence between the multiplicative and additive
forms, it is
the choice of hyperprior which affects inferences about the intercept, depending upon whether
the $E(u_i ) = 1$ or $E(\varepsilon _i ) =0$ constraint is imposed through the hyperprior.
Bayesian
applications of negative binomial modeling as well as hierarchical
PG and PLN modeling can be
found in Ntzoufras (2009) and in the references therein.
Bayesian literature on PIG modeling is limited to the study of Font et al. (2013),
which emphasizes on a marginal, zero-truncated form of the model specifically suited for linguistic analysis.

In what follows, we present the hierarchical and marginal forms and properties of the three models,
with emphasis placed on the PIG.
Markov chain Monte Carlo (MCMC) sampling is based on a posterior factorization which is not common,
but is particularly convenient in our context given the large data size.
Specifically, if we denote by $\omega$ the hyper-parameter of the mixing prior of $\bf u$,
then the joint posterior is $p(
{\boldsymbol \mu}, {\bf u}, \omega | {\bf y}) = p({\bf u} | {\boldsymbol \mu},
\omega, {\bf{y}}) p({\boldsymbol \mu}, \omega | {\bf y})$. Thus, for
hierarchical inference one can use the marginal likelihood for sampling
from $p({\boldsymbol \mu}, \omega | {\bf y})$  and generate $\bf u$
subsequently from $p({\bf u} | {\boldsymbol \mu}, \omega, {\bf y})$. As
illustrated next, this is straightforward for the PG and PIG models.

\subsection{The PG model}

For the PG model we make the following likelihood and prior
assumptions;
\[
\begin{array}{l}
 y_i \vert {\boldsymbol {\beta }},u_i \sim Poisson(\exp ({\boldsymbol {\beta }}^T{\rm{\bf x}}_i) u_i ), \\
 {\boldsymbol {\beta }}\sim {\rm {\bf N}}_{p + 1} ({\rm {\bf 0}},{\rm {\bf\Sigma }}_{\boldsymbol {\beta }} )\mbox{ with }{\rm {\bf \Sigma
}}_{\boldsymbol {\beta }} = n({\rm {\bf X}}^{\rm {\bf '}}{\rm {\bf X}})^{ - 1},\\
 u_i \sim Gamma(\theta ,\theta )\mbox{ and} \\
 \theta \sim Gamma(a,a)\mbox{ with }a = 10^{ - 3}.
 \end{array}
\]
For the multivariate normal prior of the regression parameters we
adopt the $g$-prior structure (Zellner, 1986), analogue to
the benchmark prior discussed in Fern\'{a}ndez et al. (2001) for normal linear models.
The same unit-information multivariate prior
is also adopted for the PLN and PIG models. The gamma prior for
$u_i $ is defined in terms of shape and rate parameters which both
equal $\theta $, so that $E(u_i ) = 1$ and $Var(u_i ) = \theta ^{
- 1}$. The gamma hyperprior for dispersion parameter $\theta $,
with shape and rate equal to 0.001, is a commonly used diffuse prior
(Ntzoufras, 2009). The joint posterior distribution of all
parameters is $p({\boldsymbol {\beta }},{\rm {\bf u}},\theta \vert
{\rm {\bf y}}) \propto p({\rm {\bf y}}\vert {\boldsymbol {\bf \beta
}},{\rm {\bf u}})p({\boldsymbol {\beta }})p({\rm {\bf u}}\vert \theta
)p(\theta )$.
The only full conditional which has a known form is that of the
random effects which is a gamma distribution, namely $u_i \vert
{\boldsymbol {\beta }},\theta ,y_i \sim Gamma(y_i + \theta ,\mu_i + \theta )$ (Gelman and Hill,
2006). Therefore, MCMC for the hierarchical model would require a Metropolis-within-Gibbs type of algorithm
with Metropolis steps for the joint conditional of ${\boldsymbol {\bf
\beta }},{\rm {\bf \theta }}\vert {\rm {\bf u}},{\rm {\bf y}}$ or
for the conditionals of ${\boldsymbol {\beta }}\vert {\rm {\bf
u}},{\rm {\bf y}}$ and $\theta \vert {\rm {\bf u}},{\rm {\bf y}}$.
Alternatively, adaptive rejection sampling can also be used.

Integration over \textbf{u} leads to a negative binomial marginal
likelihood, i.e. $y_i \vert {\boldsymbol {\beta }},\theta \sim
NB(\exp ({\boldsymbol {\beta }}^T{\rm{\bf x}}_i),\theta )$. Under this
parameterization the marginal mean and variance are given by
$E({\rm {\bf y}}\vert {\boldsymbol {\beta }}) =\exp({{\rm {\bf X}}{\boldsymbol {\beta }}})$
and
$
Var({\rm {\bf y}}\vert {\boldsymbol {\bf \beta
}},\theta ) = \exp({{\rm {\bf X}}{\boldsymbol {\beta }}}) + {\exp({{\rm {\bf X}}{\boldsymbol {\beta }}})}^2\theta ^{ - 1}
$,
with the variance being a quadratic function of the mean. The
posterior distribution now is $p({\boldsymbol {\beta }},\theta \vert
{\rm {\bf y}}) \propto p({\rm {\bf y}}\vert {\boldsymbol {\bf \beta
}},\theta )p({\boldsymbol {\beta }})p({\rm {\bf \theta }})$, which
leads to expression

\begin{equation}
\begin{aligned}
p({\boldsymbol {\beta }},\theta \vert {\rm {\bf y}}) \propto &
\exp { \left(
{\bf y}^T {\bf X} {\boldsymbol \beta}
- \frac{1}{2}{\boldsymbol {\bf\beta }}^T{\rm {\bf \Sigma }}_{\boldsymbol {\beta }}^{ - 1} {\boldsymbol {\bf\beta }}
- a\theta
\right)}\theta^{n\theta +a} {\Gamma (\theta )}^{ -
n} \\ &
\times
\prod\limits_i {\left[ {\Gamma \left(y_i + \theta \right)
{\left({\exp ({\boldsymbol {\beta }}^T{\rm{\bf x}}_i) + \theta } \right )^{-(y_i+ \theta) }
}} \right]}
.
\end{aligned}
\end{equation}

The Metropolis-Hastings (M-H) algorithm is used to sample from the
joint posterior of ${\boldsymbol {\beta }},\theta \vert {\rm {\bf
y}}$. Once $M$ posterior draws of ${\boldsymbol {\beta }}$ and $\theta
$ are available, predictive inference from the hierarchical
structure of the model is straightforward; we generate first ${\rm
{\bf u}}^{(m)}\sim Gamma({\rm {\bf y}} + \theta ^{(m)},\exp ({\bf X}{\boldsymbol {\beta }}^{(m)}) + \theta ^{(m)})$ and then
${\rm {\bf y}}^{pred(m)}\sim Poisson(\exp ({\bf X}{\boldsymbol {\beta }}^{(m)}){\rm {\bf u}}^{(m)})$ for $m =
1,2,...,M$. As shown in sections 4.4 and 4.5, predictions are used in posterior predictive checks and also for quantifying input-uncertainty in deterministic traffic-assignment modeling.

 It is worth noting that recent developments (Martins
and Rue, 2013) extend the initial INLA framework (Rue et al.,
2009) to applications on near-Gaussian latent models. Therefore,
we also consider INLA as an alternative to MCMC for the PG model;
a comparison is presented in Section 4.1.

\subsection{The PLN model}

The assumptions  are the following;
\[
\begin{array}{l}
 y_i \vert {\boldsymbol {\beta }},u_i \sim Poisson(\exp ({\boldsymbol {\beta }}^T{\rm{\bf x}}_i) u_i ), \\
 {\boldsymbol {\beta }}\sim {\rm {\bf N}}_{p + 1} ({\rm {\bf 0}},{\rm {\bf
\Sigma }}_{\boldsymbol {\beta }} )\mbox{ with }{\rm {\bf \Sigma
}}_{\boldsymbol {\beta }} = n({\rm {\bf X}}^T{\rm {\bf X}})^{ -
1},
\\
 u_i \sim LN( - \sigma^2 / 2,\sigma^2 )\mbox{ and} \\
 \sigma ^2 \sim InvGamma(a,a)\mbox{ with }a = 10^{ - 3}. \\
 \end{array}
\]
Following the formulation of Lee and Nelder (2004) for scale
identifiability, the prior distribution of $u_i $ has location
parameter equal to $ - \sigma^2 / 2$ and scale $\sigma^2 $,
and so $E(u_i ) = 1$ and $Var(u_i ) = {e^{\sigma^2 } -
1}$. The inverse gamma hyperprior for $\sigma ^2 $ is the common
option for this model (Ntzoufras, 2009); for $a = 10^{ - 3}$ the
distribution of $\sigma ^{ - 2} $ is a diffuse gamma. The joint
posterior distribution is $p({\boldsymbol {\beta }},{\rm {\bf
u}},\sigma^2 \vert {\rm {\bf y}}) \propto p({\rm {\bf y}}\vert
{\boldsymbol {\beta }},{\rm {\bf u}})p({\boldsymbol {\beta }})p({\rm {\bf
u}}\vert \sigma^2 )p(\sigma^2 )$.
In this case, none of the full conditional distributions are of
known form. MCMC sampling for the hierarchical PLN model is in
general more convenient in its GLMM form where the full
conditional distribution of $\sigma^2 $ is again an inverse
gamma distribution, namely
$\sigma^2 \vert {\rm {\bf u}},{\rm
{\bf y}}\sim InvGamma({a + n / 2,a + \sum_i{(
{\log u_i })^2} / 2}). $
Thus, in the additive case sampling from the
conditionals of ${\boldsymbol {\beta }}$ and \textbf{u} is possible
with Metropolis steps
or rejection-sampling.
Note that in the GLMM form
the corresponding prior for $u_i$ must be specified as $  LN( 0,\sigma^2 )$.

In the PLN model the marginal likelihood $p({\rm {\bf
y}}\vert {\boldsymbol {\beta }},\sigma^2 )$ is not known
analytically, nevertheless the mean and variance of the PLN
distribution are available and given by $E({\rm {\bf y}}\vert {\boldsymbol
{\bf \beta }}) = \exp({{\rm {\bf X}}{\boldsymbol {\beta }}})$ and $Var({\rm {\bf y}}\vert {\boldsymbol {\beta }},\sigma^2 )
=\exp({{\rm {\bf X}}{\boldsymbol {\beta }}}) + {\exp({{\rm {\bf X}}{\boldsymbol {\beta }}})}^2( {\exp(\sigma^2) - 1}
)$. As with the PG model, the variance is a quadratic function of the
mean. The joint posterior density is $p({\boldsymbol {\beta }},\sigma^2 \vert {\rm {\bf y}}) \propto p({\rm {\bf y}}\vert {\boldsymbol {\bf
\beta }},\sigma^2 )p({\boldsymbol {\beta }})p(\sigma^2 )$,
namely

\begin{equation}
\begin{aligned}
p({\boldsymbol {\beta }},\sigma^2 \vert {\rm {\bf y}})
\propto &
\prod\limits_i {\left[ {\int { {\exp \left ({y_i{\boldsymbol {\bf \beta
}}^T{\rm {\bf x}}_i  - \exp{{\boldsymbol {\bf \beta
}}^T{\rm {\bf x}}_i }u_i - \frac{(\log u_i + \sigma^2 / 2)^2}{2\sigma^2 }} \right ) } u_i^{ y_i- 1}d u_i } } \right]} \\
&\times \exp{\left ( - \frac{1}{2}{\boldsymbol {\bf
\beta }}^T{\rm {\bf \Sigma }}_{\boldsymbol {\beta }}^{ - 1} {\boldsymbol {\bf
\beta }} - a / \sigma^2 \right ) } \left( {\sigma^2 } \right)^{ - (n/2 + a + 1)}.
\end{aligned}
\end{equation}

We employ M-H simulation in order to sample from the joint
posterior density of ${\boldsymbol {\beta }}$ and $\sigma^2 $. The
integral appearing in the un-normalized posterior can be evaluated through
numerical integration, e.g. with Gauss-Hermite quadrature
which is also frequently employed in frequentist
practice for marginal estimation. Another alternative examined in
this study is Monte Carlo (MC) integration from the lognormal
prior of the random effect vector \textbf{u }within the Metropolis
kernel. That is, for a given M-H iteration $t$ and draws ${\boldsymbol {\bf
\beta }}^{(t)},\sigma^{2(t)} $, the above integral can be
evaluated by generating first $L$ draws $\{ {u_i^{(t,l)}
,\mbox{ }l = 1,2,...,L} \}$ from $u_i^{(t,l)} \sim LN( -
\sigma^{2(t)} / 2,\sigma^{2(t)} )$ and then by calculating
the marginal probability as $p(y_i \vert {\boldsymbol {\bf \beta
}}^{(t)},\sigma^{2(t)} ) = L^{ - 1}\sum_l {p(y_i \vert
{\boldsymbol {\beta }}^{(t)},u_i^{(t,l)} )} $.

A potentially efficient alternative to MCMC
approaches for the PLN model is the INLA framework introduced in
Rue et al. (2009). The INLA approach covers the family of Gaussian
Markov random fields (GMRF) models and is based on efficient
approximating schemes for the marginal posterior distributions.
The PLN model is included in the family of GMRF models as the
random effects are normally distributed on additive scale. In Section 4.1 we compare INLA to MCMC.

\subsection{ The PIG model}

For the hierarchical PIG we adopt the following
assumptions;
\[
\begin{array}{l}
 y_i \vert {\boldsymbol {\beta }},u_i \sim Poisson(\exp ({\boldsymbol {\beta }}^T{\rm{\bf x}}_i) u_i ), \\
 {\boldsymbol {\beta }}\sim {\rm {\bf N}}_{p + 1} ({\rm {\bf 0}},{\rm {\bf
\Sigma }}_{\boldsymbol {\beta }} )\mbox{ with }{\rm {\bf \Sigma
}}_{\boldsymbol {\beta }} = n({\rm {\bf X}}^T{\rm {\bf X}})^{ -
1},
\\
 u_i \sim IG(1,\zeta )\mbox{ and} \\
 \zeta \sim Gamma(a,a)\mbox{ with }a = 10^{ - 3}. \\
 \end{array}
\]
The initial parameterization of Holla
(1967) is used for the IG prior, with mean $\mu $ and shape $\zeta $, specifically
\begin{equation}
p(u_i \vert \mu ,\zeta ) = \left( {\frac{\zeta }{2\pi u_i^3 }}
\right)^{1 / 2}\exp \left({ - \frac{\zeta (u_i - \mu )^2}{2\mu ^2u_i }} \right).
\end{equation}
For $\mu = 1$ we have that a-priori $E(u_i ) = 1$ and $Var(u_i ) =
\zeta ^{ - 1}$. The IG distribution is a special case of the three
parameter generalized inverse Gaussian (GIG) distribution which is
generally conjugate to the family of exponential distributionsis
and is studied in detail in Jorgensen (1982). The p.d.f. of a
GIG($\lambda ,\psi ,\chi )$ distribution with parameters $\lambda
\in {\rm R}$, $\chi ,\psi > 0$ is given by
\begin{equation}
f(x) = \frac{(\psi / \chi )^{\lambda / 2}}{2K_\lambda (\sqrt {\psi
\chi } )}x^{\lambda - 1}\exp \left[{ - \frac{1}{2}\left( {\psi x + \chi x^{
- 1}} \right)} \right],
\end{equation}
where $K_\lambda $ is the modified Bessel function of the third
kind with order $\lambda .$ The IG distribution arises for
$\lambda = - 1 / 2$. Interestingly, the gamma distribution is also
a special case of the GIG distribution for $\chi = 0$. For shape
parameter $\zeta $ we adopt the usual gamma hyperprior, similarly
to the PG model. The posterior distribution now is $p({\boldsymbol {\bf
\beta }},{\rm {\bf u}},\zeta \vert {\rm {\bf y}}) \propto p({\rm
{\bf y}}\vert {\boldsymbol {\beta }},{\rm {\bf u}})p({\boldsymbol {\bf \beta
}})p({\rm {\bf u}}\vert \zeta )p(\zeta )$
and it can be easily shown that the full conditionals of \textbf{u}
and $\zeta $ are known distributions, namely $u_i \vert \boldsymbol{\beta}, \zeta \sim GIG(y_i - 1
/ 2,2\exp({{\boldsymbol {\beta }}^T{\rm {\bf x}}_i }) + \zeta ,\zeta )$
and $\zeta \vert {\bf u} \sim Gamma(a + n / 2,a + \sum_i {(u_i - 1)^2 /
2u_i } )$.
Athreya (1986) was the first to notice the
specific conjugate relationship between the IG and Poisson
distribution; see also Karlis (2001). Regarding simulation from the
GIG distribution, random generators are readily available (e.g.
Dagpunar, 1988).
Thus, the hierarchical PIG model is actually simpler
in terms of MCMC in comparison to the PG and PLN models, since all
that is needed is a M-H step or rejection-sampling algorithm for the
conditional of ${\boldsymbol {\beta }}$.

Marginally we have that $y_i \vert {\boldsymbol {\beta }},\zeta \sim
PIG(\exp({{\boldsymbol {\beta }}^T{\rm {\bf x}}_i }),\zeta )$ for $i =
1,2,...,n$ with p.d.f. given by
\begin{equation}
\begin{aligned}
p(y_i \vert {\boldsymbol {\beta }},\zeta ) = & K_{y_i - 1 / 2} \left(
{\sqrt {2\zeta \phi({\bf x}_i)} } \right)\left( {\frac{2\zeta
}{\pi }} \right)^{1 / 2}\frac{\exp \left({\zeta / \exp({{\boldsymbol {\beta }}^T
{\bf x}_i })}\right)}{y_i !} \\
\times &
\left( {2 \phi({\bf x}_i) \zeta ^{ - 1}}
\right)^{1 / 2(y_i - 1 / 2)},
\end{aligned}
\end{equation}
where
\begin{equation}
\phi({\bf x}_i) = \left(1 + \frac{\zeta }{2\exp({{\boldsymbol \beta }^T
{\bf x}_i })} \right).
\end{equation}
The marginal mean and variance are
$E({\rm {\bf y}}\vert {\boldsymbol {\beta }}) = \exp({{\rm {\bf X}}{\boldsymbol {\beta }}})$ and $Var({\rm {\bf y}}\vert {\boldsymbol {\bf \beta
}},\zeta ) = \exp({{\rm {\bf X}}{\boldsymbol {\beta }}}) + (\exp({{\rm {\bf X}}{\boldsymbol {\beta }}}))^3\zeta ^{ - 1}$.
The variance is thus a cubic function of the mean in the PIG
model, allowing for greater overdispersion. The posterior distribution $p({\boldsymbol {\bf \beta
}},\zeta \vert {\rm {\bf y}}) \propto p({\rm {\bf y}}\vert {\boldsymbol
{\bf \beta }},\zeta )p({\boldsymbol {\beta }})p(\zeta )$ can be expressed as
\begin{equation}
\begin{aligned}
p({\boldsymbol {\beta }},\zeta \vert {\rm {\bf y}}) \propto&
\prod\limits_i {\left[ {K_{y_i - 1 / 2} \left(
{\sqrt {2\zeta \phi({\bf x}_i)} } \right) \exp{ \left( \zeta / \exp{({\boldsymbol {\bf
\beta }}^T{\rm {\bf x}}_i }) \right)}\left( {2\phi({\bf x}_i)\zeta^{ - 1}}
\right)^{1 / 2(y_i - 1 / 2)}} \right]}
\\
& \times \exp{ \left(- \frac{1}{2}{\boldsymbol {\beta }}^T{\rm {\bf \Sigma
}}_{\boldsymbol {\beta }}^{ - 1} {\boldsymbol {\beta }}- a\zeta\right)} \zeta ^{n/2+a - 1}.
\end{aligned}
\end{equation}
Samples from the posterior of ${\boldsymbol {\beta }}$ and $\zeta $
can be obtained through M-H simulation from the joint posterior. As
with the PG model, when $M$ posterior draws of ${\boldsymbol {\beta }}$
and $\zeta $ are available, predictive inference from the
hierarchical structure of the PIG model is possible by generating
first ${\rm {\bf u}}^{(m)}\sim GIG({\rm {\bf y}} - 1 / 2,2\exp ({\bf X}{\boldsymbol {\beta }}^{(m)}) + \zeta ^{(m)},\zeta
^{(m)})$ and then ${\rm {\bf y}}^{pred(m)}\sim Poisson(\exp ({\bf X}{\boldsymbol {\beta }}^{(m)}){\rm {\bf u}}^{(m)})$ for $m
= 1,2,...,M$.

\section{Data}
\subsection{The OD matrix and the transport network of Flanders}

The OD matrix was derived from the 2001 Belgian census
study and contains information about the departure and arrival
locations for work and school related trips of the approximately
10 million Belgian residents. The recorded work/school trips refer
to a normal weekday for all possible travel modes and are
one-directional, from zone of origin to zone of destination. The
study area is not the entire country of Belgium, but the
northern, Dutch-speaking region of Flanders which roughly accounts
for 60{\%} of the total population and 44{\%} of the country's
surface area. From an administrative viewpoint Flanders is divided
into 5 provinces, 22 arrondissements, 52 districts, 103 cantons
and 308 municipalities. Our analysis is implemented on the
municipal level at which the OD matrix contains 94864 cells.

The OD flows on municipality zonal level are sparsely
distributed and extremely overdispersed with large outlying
observations. Approximately 63{\%} of the observations are
zero-valued, with an overall mean of 38.47 and a standard
deviation of 960.47. All of the zero-valued observations belong to
inter-zonal flows off the main diagonal. The mean and standard
deviation of inter-zonal flows are equal to 18.48 and 156.67,
respectively. The maximum value is observed in the diagonal cell
which corresponds to the intra-zonal flows occurring in Antwerp --
the largest Flemish municipality -- and is equal to 211681. In general, the
majority of trips correspond to intra-zonal flows with the counts
on the main diagonal accounting for approximately 51{\%} of the
total number of trips. The mean for intra-zonal flows is 6064.32,
while the standard deviation is 15516.64.

\begin{figure}
\begin{center}
 \includegraphics[scale=0.38]{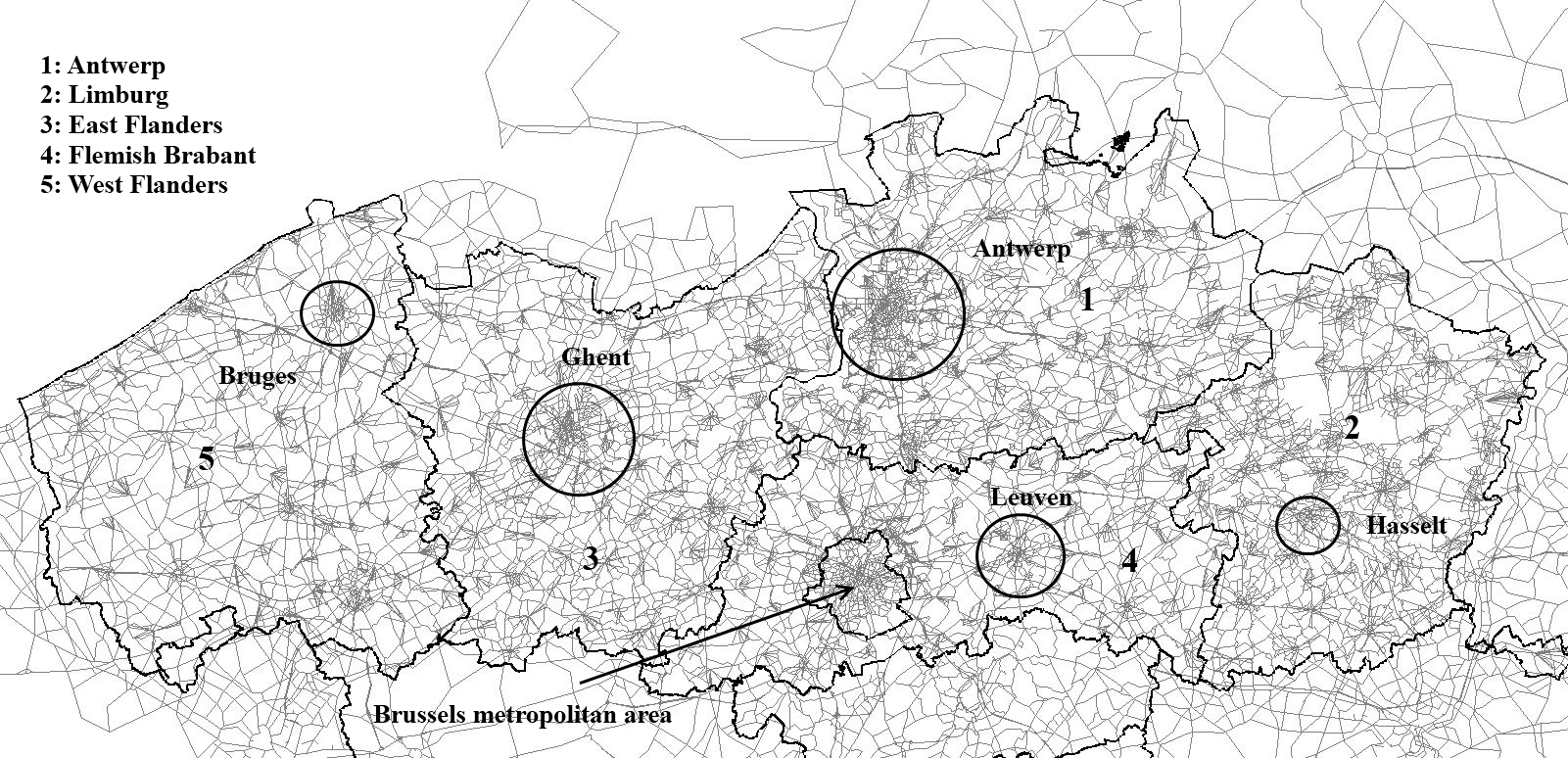}
\caption{\label{network} The road network of Flanders and the 5
Flemish provinces of Antwerp, Limburg, East Flanders, Flemish
Brabant and West Flanders with corresponding capitals; Antwerp,
Hasselt, Ghent, Leuven and Bruges.}\end{center}
\end{figure}

The road network of Flanders with the corresponding borders of the
5 Flemish provinces, Antwerp, Limburg, East Flanders, Flemish
Brabant and West Flanders, is presented in Figure \ref{network}.
The circled areas indicate the capital-municipality of each
province, the size of each circle is a relative representation of
population size. Antwerp is the most populated capital, followed
by Ghent, Leuven, Bruges and Hasselt. Brussels metropolitan area,
which is also marked in the map, is not included in the analysis
as it is a separate administrative center.
In overall, the
network runs a total length of 65296.72 kilometers and contains
97450 links which can be categorized into highways (8.58{\%}
including entrance/exit road segments), main regional roads
(15.49{\%}), small regional roads (21.1{\%}), local municipal
roads (52.91{\%}) and walk/bicycle paths (1.92{\%}).

\subsection{ Explanatory variables}

The set of explanatory variables consists of six categorical variables
and twelve discrete/continuous variables. The first five categorical variables capture the
effects of intra-zonal flows measured in differences of 100 trips. Thus,
these dummies take the value 100 if the trips
are intra-zonal in municipalities (DM), cantons (DC), districts
(DD) arrondissements (DA) provinces (DP) and 0 otherwise. These
predictors capture individual effects; for
instance, for intra-zonal flows in the main diagonal the
municipality predictor DM will equal 100, whereas DC, DD, DA and DP will
equal 0. The sixth categorical predictor (DE) is associated with the effect of
higher education institutes in destination zones; it takes the
value of 1 if the destination zone supports a college and/or a
university and 0 otherwise. The set of covariates includes four
discrete-valued variables which contain the total number of
neighboring municipalities on canton (MC), district (MD),
arrondissement (MA) and province (MP) levels for each
corresponding OD pair. The rest of the covariates are
continuous. Specifically, we include employment rate (ER),
population density (PD; thousands inhabitants per square km),
relative length of road networks (RL; road length in km's per
surface area in square km's), perimeter length (PL) in km's, car
ownership ratio (CR), yearly traffic in highways (HT) and in
provincial/municipal roads (PMT) in km's, and finally distance (D)
in km's. All covariates are used in logarithmic scale. Distance,
of course, is zero for intra-zonal municipality flows and in order
to use the logarithm it is set equal to 0.1, a value which for
most practical purposes refers to negligible distance (100
meters). Furthermore, due to the particularity of the OD problem
variables ER, PD, RL, PL, CR, HT and PMT come in pairs, i.e. each
is used twice, one time for the origin-zone and one time for the
destination-zone. The arguments for employing the
continuous variables in pairs are the following; a) preliminary
research revealed that it is better to use information for origin
and destination zones separately rather than average, for
instance, between origin and destination zones,
b) having separate parameters estimates for origin and destination
zones allows for elementary comparison with trip-production/attraction,
c) using pairs on logarithmic scale and
including distance provides an alternative interpretation of the
Poisson mixture log-linear models as stochastic gravity,
direct-demand models.

Most of the continuous variables were transformed to ratios
relative to populations or surface areas. The specific
transformations were chosen in order to maintain reasonable
interpretations, but also in order to solve multicollinearity
problems which were evidently present in raw variables.
Analysis based on variance inflation factors
(VIF) indicated no serious multicollinearity problems for the
transformed variables with the highest VIF value being equal to
3.877.

\section{Results}

We start this section with a comparison between MCMC and INLA estimates for the PLN and PG models on an OD matrix of smaller scale. The full analysis for entire Flanders, including posterior and predictive inference based on MCMC, is presented next. Details concerning M-H implementation are presented in Appendix A.

\subsection{Comparing MCMC and INLA}

The comparison presented here concerns a 10 by 10 OD matrix
containing the flows between the 10 largest (in terms of
population)  Flemish municipalities. The rationale in choosing a
smaller OD matrix is to evaluate how well can INLA approximate
marginal posterior distributions under relatively small samples.
The categorical predictors are not meaningful to use in this case,
therefore, we use only employment rate, population density, length
of road networks, highway traffic, provincial/municipal traffic
and distance as covariates.

The reasons for considering a smaller OD are the following. First, INLA is based on the assumption of conditional independence for the Gaussian latent random field which means that the inverse covariance matrix is sparsely distributed allowing for fast and efficient Cholesky decomposition. In general, the assumption of conditional independence becomes stronger as the dimensionality of the random field increases. Therefore, we find it interesting to evaluate INLA on a smaller random field (i.e. 100 random effects, instead of 94864, plus the regression parameters).
Second, INLA estimates the marginal posteriors either through Gaussian or through Laplace approximations based on Taylor's expansions around the posterior modes. Such approximations generally perform well when the sample size is large and the marginal posteriors are usually well centered around the posterior mode. Thus, we are interested in testing INLA on a smaller subset of the data.

For comparison purposes, we change the prior assumption for the
intercept and the regression parameters. Specifically, instead of
using the unit-information prior $ {\boldsymbol {\beta }}\sim {\rm
{\bf N}}_{12} ({\rm {\bf 0}},{\rm {\bf\Sigma }}_{\boldsymbol
{\beta }} )\mbox{ with }{\rm {\bf \Sigma}}_{\boldsymbol {\beta }}
= n({\rm {\bf X}}^T{\rm {\bf X}})^{ -1}$, we assume independent
normal priors with a large variance, namely $ {\boldsymbol {\beta
}}\sim {\rm {\bf N}}_{12} ({\rm {\bf 0}},{\rm {{{\bf\ I}_{12}
\sigma^2 }}}) \mbox{ with } \sigma^2=10^3$. For fitting the PLN
model through INLA, we used the R-INLA package (www.r-inla.org).
We consider the 3 INLA approximating strategies, namely the
Gaussian, the simplified Laplace and the Laplace approximations
for marginal posterior distributions; see Rue et al. (2009) for
details. In addition, as mentioned in Section 2.1, recent developments extend INLA to
near-Gaussian latent models (Martins and Rue, 2013). The gamma
prior is included in the available options of the R-INLA package,
which gives us the opportunity to compare MCMC and INLA for the PG
model as well.

\begin{table}
\centering{}%
\begin{tabular}{lcccc}
\hline
\multicolumn{5}{c}{\textbf{PLN estimates}}\tabularnewline
\hline
\multirow{2}{*}{\textbf{Parameter}} & \multirow{2}{*}{\textbf{M-H}} & \multicolumn{3}{c}{\textbf{INLA}}\tabularnewline
 &  & \textbf{Gaussian} & \textbf{Simplified Laplace} & \textbf{Laplace}\tabularnewline
\hline
$\beta_{0}$ intercept & -1.031 (2.288) & -1.022 (2.623) & -1.061 (2.623) & -1.061 (2.623)\tabularnewline
$\beta_{1}$ ER (o) & 0.763 (0.865) & 0.749 (0.985) & 0.755 (0.985) & 0.755 (0.985)\tabularnewline
$\beta_{2}$ ER (d) & 1.826 (0.883) & 1.802 (0.979) & 1.824 (0.979) & 1.824 (0.979)\tabularnewline
$\beta_{3}$ PD (o) & 0.406 (0.420) & 0.391 (0.476) & 0.401 (0.476) & 0.401 (0.476)\tabularnewline
$\beta_{4}$ PD (d) & 1.414 (0.425) & 1.401 (0.477) & 1.420 (0.477) & 1.420 (0.477)\tabularnewline
$\beta_{5}$ RL (o) & 0.697 (0.779) & 0.684 (0.892) & 0.689 (0.891) & 0.689 (0.891)\tabularnewline
$\beta_{6}$ RL (d) & -0.060 (0.805) & -0.048 (0.895) & -0.057 (0.894) & -0.057 (0.894)\tabularnewline
$\beta_{7}$ HT (o) & -0.297 (0.154) & -0.291 (0.180) & -0.291 (0.180) & -0.291 (0.178)\tabularnewline
$\beta_{8}$ HT (d) & 0.194 (0.156) & 0.181 (0.180) & 0.187 (0.180) & 0.187 (0.180)\tabularnewline
$\beta_{9}$ PMT (o) & 0.891 (0.225) & 0.897 (0.260) & 0.901 (0.260) & 0.901 (0.260)\tabularnewline
$\beta_{10}$PMT (d) & 0.886 (0.220) & 0.889 (0.260) & 0.890 (0.260) & 0.890 (0.259)\tabularnewline
$\beta_{11}$D & -1.129 (0.048) & -1.131 (0.057) & -1.135 (0.057) & -1.135 (0.057)\tabularnewline
$\tau (1/\sigma^2)$ & 0.989 (0.139) & 0.906 (0.139) & 0.906 (0.139) & 0.906 (0.139)\tabularnewline
\hline
\end{tabular}\caption{\label{PLN_INLA}Posterior means and standard deviations (in parentheses) from a M-H
sample of 20000 draws and from the 3 INLA approaches for the PLN model;
(o) refers to origin effects, (d) to destination effects.}
\end{table}

\begin{table}
\centering{}%
\begin{tabular}{lcccc}
\hline
\multicolumn{5}{c}{\textbf{PG estimates}}\tabularnewline
\hline
\multirow{2}{*}{\textbf{Parameter}} & \multirow{2}{*}{\textbf{M-H}} & \multicolumn{3}{c}{\textbf{INLA}}\tabularnewline
 &  & \textbf{Gaussian} & \textbf{Simplified Laplace} & \textbf{Laplace}\tabularnewline
\hline
$\beta_{0}$ intercept & -2.013 (2.488) & -2.147 (2.496) & -2.087 (2.496) & -2.052 (2.484)\tabularnewline
$\beta_{1}$ ER (o) & 1.081 (0.925) & 1.039 (0.918) & 1.016 (0.918) & 1.058 (0.913)\tabularnewline
$\beta_{2}$ ER (d) & 1.542 (0.895) & 1.486 (0.901) & 1.473 0.901) & 1.515 (0.897)\tabularnewline
$\beta_{3}$ PD (o) & 0.507 (0.428) & 0.512 (0.441) & 0.526 (0.441) & 0.513 (0.439)\tabularnewline
$\beta_{4}$ PD (d) & 1.282 (0.440) & 1.271 (0.434) & 1.289 (0.434) & 1.272 (0.432)\tabularnewline
$\beta_{5}$ RL (o) & 0.765 (0.785) & 0.718 (0.781) & 0.757 (0.781) & 0.763 (0.776)\tabularnewline
$\beta_{6}$ RL (d) & 0.078 (0.823) & -0.033 (0.810) & 0.012 (0.810) & 0.022 (0.805)\tabularnewline
$\beta_{7}$ HT (o) & -0.431 (0.180) & -0.434 (0.177) & -0.436 (0.177) & -0.440 (0.176)\tabularnewline
$\beta_{8}$ HT (d) & 0.056 (0.188) & 0.077 (0.171) & 0.073 (0.171 ) & 0.068 (0.170)\tabularnewline
$\beta_{9}$ PMT (o) & 1.082 (0.246) & 1.099 (0.243) & 1.087 (0.243) & 1.095 (0.242)\tabularnewline
$\beta_{10}$PMT (d) & 1.163 (0.232) & 1.166 (0.232) & 1.156 (0.232) & 1.161 (0.231)\tabularnewline
$\beta_{11}$D & -1.184 (0.079) & -1.168 (0.081) & -1.168 (0.081) & -1.181 (0.082)\tabularnewline
$\theta$ & 1.070 (0.123) & 1.070 (0.142) & 1.070 (0.142) & 1.070 (0.142)\tabularnewline
\hline
\end{tabular}\caption{\label{PG_INLA} Posterior means and standard deviations (in parentheses) from a M-H
sample of 20000 draws and from the 3 INLA approaches for the PG model;
(o) refers to origin effects, (d) to destination effects.}
\end{table}

\begin{figure}
\begin{center}
 \includegraphics[scale=0.45]{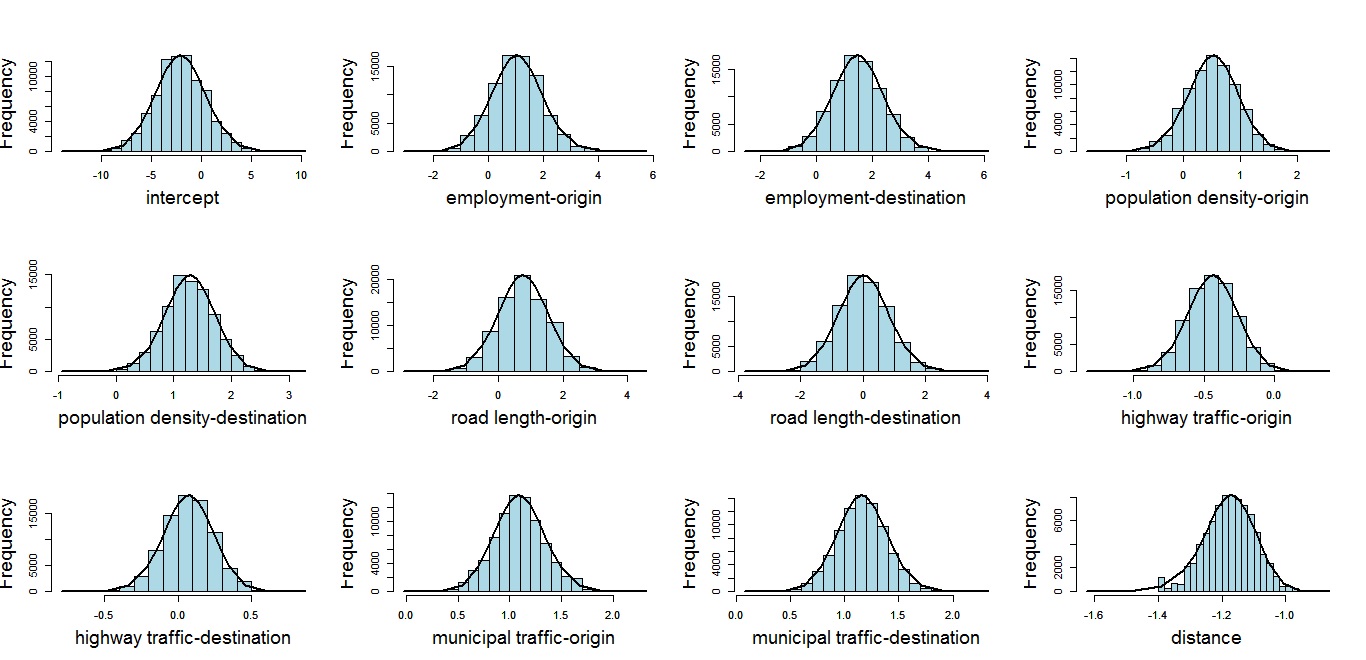}
\caption{\label{PG_marginals} Histograms of PG regression parameters from 20000 posterior draws and estimates of the posterior marginals (black lines) from INLA based on the Laplace approximation}\end{center}
\end{figure}

Posterior means and standard deviations from M-H samples of 20000
draws and from the 3 INLA approximations for the PLN and PG models
are presented in Table \ref{PLN_INLA} and Table \ref{PG_INLA},
respectively. The PLN intercept corresponds to that of the
additive model formulation. As seen, the posterior PLN means from
MCMC and INLA agree in general, especially under the simplified
Laplace and Laplace approximations. The precision estimates
slightly differ; the MCMC estimate is closer to the corresponding
ML estimate which is 1.036. Also, the standard deviations from
MCMC are overall lower. Concerning the PG model, the INLA Laplace
seems to provide more accurate estimates which are closer to the
MCMC estimates in comparison to the Gaussian and simplified
Laplace approaches. Standard deviations are more or less the same
for this model. Marginal posterior distribution estimates for the
regressors of the PG model under the Laplace approximation are
shown in Figure \ref{PG_marginals}; as seen the estimates
approximate particularly well the histograms derived from MCMC.
The corresponding figures for the PLN model (not presented here)
are equivalent.

In conclusion, we find that INLA provides a fast and efficient
alternative to MCMC under specific prior assumptions, which makes
it a potentially promising tool for OD modeling on large scale
networks. It is worth mentioning that INLA runtimes were 2.64
sec's for the PG model and 2.14 sec's for the PLN model. On the
other hand, 21000 M-H iterations (we used the first 1000 as
burn-in) required 14.82 sec's for the PG model and approximately
21 min's for the PLN model due to additional numerical integration
within MCMC. Thus, INLA is particularly useful for the PLN model.
Arguably, in medium-sized examples like this one, using MCMC for
the hierarchical data-augmented PLN model could be more efficient
than MCMC with numerical integration for the marginal likelihood.
We tested this using WinBUGS (Spiegelhalter et al., 2003),
nevertheless, the sampler failed to converge even after 200000
iterations. All computations were performed in a  standard 64-bit
laptop with 2.20 GHz CPU and 4 GB of RAM using R version 3.0.1.
The R code used for INLA and the subset of the OD data are
provided as supporting material.

Despite these advantages, we find that the R-INLA package is still restrictive with respect to certain aspects and requires further development which will allow for more general modeling frameworks. In particular, the package does not yet fully support multivariate prior assumptions such as $g$-prior structures (e.g. Zellner, 1986) for regression coefficients. Moreover, it would be interesting to include further distributional options covering near-Gaussian latent fields, such as the inverse-Gaussian prior considered in this paper.

\subsection{Posterior inference for Flanders}

Posterior means and 95{\%} credible intervals based on 4000
posterior draws are presented in Table \ref{mainresult}. The
corresponding INLA Gaussian estimates for the PG and PLN models
are presented in Appendix B. The INLA estimates are slightly
different due to the different prior assumption, namely $
{\boldsymbol {\beta }}\sim {\rm {\bf N}}_{12} ({\rm {\bf 0}},{\rm
{{{\bf\ I}_{12} \sigma^2 }}}) \mbox{ with } \sigma^2=10^3$.
Nevertheless, the overall conclusions discussed next are also
supported by the INLA estimates. Details of M-H implementation and
a comparison of MCMC and INLA runtimes can be found in Appendix A.
In general, the posterior means of the PLN and PIG models are more
similar. For instance, parameters $\beta _0 $, $\beta _6 $, $\beta
_9 $, $\beta _{10} $, $\beta _{14} $ and $\beta _{18} $ of the PG
model are substantially different from the corresponding estimates
of the other two models, especially the intercept estimate. On the
other hand, parameters $\beta _{11} $, $\beta _{12} $ and $\beta
_{20} $ differ across models.

The parameters $\beta _1 $ to $\beta _5 $ of the categorical variables are all
positive except of the last parameter for intra-zonal municipality
trips. The positive effects of $\beta _1 $ to $\beta _4 $ are to
be expected, since the OD flows are generally larger in diagonal
blocks of cells of the OD matrix corresponding to intra-zonal
flows for the various administrative levels. The negative sign of
$\beta _5 $ is not expected but it might be explained as simply
counterbalancing the absence of the strong negative effect of
distance which is set almost equal to zero for intra-zonal
municipality trips. Parameter $\beta _6 $ is positive which leads to
the consistent interpretation that destination zones which support
a college or a university are more likely to attract trips than
zones without a college/university.

Parameters $\beta _7 $ to $\beta _{10} $ quantify the influence of
the total number of surrounding municipalities on the levels of
cantons, districts, arrondissements and provinces, respectively.
This effect is in general not straightforward to predict,
nevertheless the parameter estimates provide some insights. On the
small-scale level of cantons parameter $\beta _7 $ has a positive
sign, whereas on the large-scale levels of districts,
arrondissements and provinces -- where the total number of
municipalities increase and a spead-out of trips is more likely --
the corresponding parameters $\beta _8 $, $\beta _9 $ and $\beta
_{10} $ are negative. This implies that the effect changes from
positive to negative when exceeding a specific radius threshold of
distance.
Recent transportation studies discuss similar ideas such
as the \textit{neighborhood-effect} concept investigated in more detail by Sohn and Kim (2010).

Regarding the continuous variables used in pairs, the more general
explanatory variables have parameters with positive signs, namely
population density ($\beta _{13} $,$\beta _{14} )$, perimeter
length ($\beta _{17} $,$\beta _{18} )$ and
kilometers-driven in highways ($\beta _{21} $,$\beta _{22} )$ and
provincial/municipal roads ($\beta _{23} $,$\beta _{24} )$. The
uniformly positive effects for origin and destination zones do not
come as a surprise, since we would expect these four variables to
be positively correlated with trip-production (origin zones) as
well as trip-attraction (destination zones). In contrast, the
parameters of employment rate ($\beta _{11} $,$\beta _{12} )$,
relative length of road network ($\beta _{15} $,$\beta _{16} )$
and car ownership ratio ($\beta _{19} $,$\beta _{20} )$ have
opposite signs for origin and destination effects.

\begin{table}[htbp]
\begin{center}
\begin{tabular}
{ccccccc} \hline \raisebox{-1.50ex}[0cm][0cm]{\textbf{Parameter}}&
\multicolumn{2}{c}{\textbf{PG}} &
\multicolumn{2}{c}{\textbf{PLN}} &
\multicolumn{2}{c}{\textbf{PIG}}  \\
 &
Mean&  95{\%} Cr. Int.& Mean&  95{\%} Cr. Int.& Mean&
95{\%} Cr. Int. \\
\hline $\beta _0 $ & 4.027&  (3.214, 4.869)& 6.104&
(5.230, 6.989)&
6.847&
(6.028, 7.675) \\
$\beta _1 $ DP& 0.005& (0.005, 0.005)& 0.005& (0.005, 0.006)&
0.006&
(0.005, 0.006) \\
$\beta _2 $ DA& 0.007&  (0.007, 0.008)& 0.007& (0.007, 0.008)&
0.008&
(0.007, 0.008) \\
$\beta _3 $ DD& 0.008& (0.008, 0.009)& 0.008& (0.008, 0.009)&
0.009&
(0.008, 0.009) \\
 $\beta _4 $ DC& 0.008&  (0.007, 0.009)& 0.006& (0.006, 0.007)&
0.006&
(0.006, 0.007) \\
 $\beta _5 $ DM& -0.082&  (-0.083, -0.080)& -0.086& (-0.087,
-0.084)&
-0.084&
(-0.086, -0.083) \\
 $\beta _6 $ DE& 0.424&  (0.388, 0.460)& 0.535& (0.496, 0.574)&
0.536&
(0.497, 0.581) \\
 $\beta _7 $ MC& 0.473&  (0.435, 0.510)& 0.461& (0.422, 0.501)&
0.450&
(0.411, 0.490) \\
 $\beta _8 $ MD& -0.494& (-0.542, -0.445)& -0.441& (-0.491,
-0.391)&
 -0.442&
(-0.489, -0.392) \\
 $\beta _9 $ MA& -0.088&  (-0.124, -0.055)& -0.188& (-0.225,
-0.149)&
 -0.210&
(-0.249, -0.167) \\
 $\beta _{10} $ MP& -0.491&  (-0.636, -0.345)& -0.737&
 (-0.888, -0.589)&
 -0.783&
(-0.924, -0.636) \\
 $\beta _{11} $ ER(o)& -1.062&  (-1.207, -0.918)& -0.482&
(-0.629, -0.334)&
 -0.240&
(-0.383, -0.105) \\
 $\beta _{12} $ ER(d)& 0.326& (0.194, 0.462)& 0.492&
 (0.345, 0.641)&
 0.608&
(0.460, 0.759) \\
 $\beta _{13} $ PD(o)& 0.505& (0.477, 0.533)& 0.499& (0.470,
0.528)&
 0.500&
(0.474, 0.529) \\
 $\beta _{14} $ PD(d)& 0.577&  (0.548, 0.606)& 0.626&
 (0.594, 0.658)&
 0.631&
(0.595, 0.662) \\
 $\beta _{15} $ RL(o)& -0.315& (-0.359, -0.272)& -0.318&
 (-0.365, -0.272)&
 -0.334& (-0.380, -0.289) \\
 $\beta _{16} $ RL(d)& 0.280& (0.236, 0.322)& 0.267& (0.220,
0.315)&
 0.265&
(0.219, 0.310) \\
 $\beta _{17} $ PL(o)& 1.253& (1.208, 1.298)& 1.289& (1.241,
1.338)&
1.283&
(1.238, 1.327) \\
 $\beta _{18} $ PL(d)& 0.430&  (0.385, 0.475)& 0.500&
 (0.452, 0.549)&
 0.509&
(0.459, 0.559) \\
 $\beta _{19} $ CR(o)& 3.454& (3.149, 3.762)& 3.413& (3.095,
3.731)&
3.520&
(3.227, 3.831) \\
 $\beta _{20} $ CR(d)& -1.465& (-1.768, -1.180)& -1.255& (-1.577,
-0.939)&
-1.081&
(-1.386, -0.752) \\
 $\beta _{21} $ HT(o)& 0.010& (0.007, 0.014)& 0.011& (0.008,
0.015)&
0.010&
(0.007, 0.014) \\
 $\beta _{22} $ HT(d)& 0.052&  (0.049, 0.056)& 0.050&
 (0.047, 0.054)&
 0.050&
(0.046, 0.053) \\
 $\beta _{23} $ PMT(o)& 0.270&  (0.250, 0.289)& 0.278&
 (0.257, 0.299)&
 0.275&
(0.254, 0.294) \\
 $\beta _{24} $ PMT(d)& 0.869&  (0.850, 0.888)& 0.876&
 (0.853, 0.898)&
 0.870&
(0.849, 0.891) \\
 $\beta _{25} $ D& -2.906&  (-2.927 -2.885)& -2.984& (-3.007,
-2.960)&
 -2.936&
(-2.957, -2.915) \\
 $\theta $& 0.965&  (0.947, 0.983)& \multicolumn{2}{c}{-} &
\multicolumn{2}{c}{-}  \\
$\sigma^2 $& \multicolumn{2}{c}{-} & 1.065& (1.043, 1.086)&
\multicolumn{2}{c}{-}  \\
 $\zeta $& \multicolumn{2}{c}{-} & \multicolumn{2}{c}{-} &
0.377& (0.359, 0.399) \\
 AIC & \multicolumn{2}{c}{281519.3} & \multicolumn{2}{c}{279364.7}
&
\multicolumn{2}{c}{278468.9}  \\
 BIC& \multicolumn{2}{c}{281774.7} & \multicolumn{2}{c}{279620.1}
&
\multicolumn{2}{c}{278724.3}  \\
 DIC (marginal) & \multicolumn{2}{c}{281492.4} &
\multicolumn{2}{c}{279337.7} &
\multicolumn{2}{c}{278441.4}  \\
 DIC (hierachical)& \multicolumn{2}{c}{224141.4} &
\multicolumn{2}{c}{-} &
\multicolumn{2}{c}{224146.1}  \\
\hline
\end{tabular}
\caption{\label{mainresult}Posterior means and 95{\%} credible
intervals for regression and dispersion parameters and the values
of AIC, BIC, marginal DIC and hierarchical DIC. }
\end{center}
\end{table}

In transportation studies employment rate is commonly associated
with trip-attraction models (see e.g. Yao and Morikawa, 2005). In
accordance, the posterior estimate of employment rate is positive
for destination zones and negative for origin zones which leads to
the rational interpretation that zones with high employment rates
are more likely to attract trips rather than to generate trips.
The relative length of road networks is associated with the
concept of \textit{accessibility} (see e.g. Odoki et al., 2001), a concept which is present primarily in
trip-attraction studies.
In general, a larger relative length in the network, will decrease the friction of travel (e.g. distance, time) significantly, and thus increase  accessibility.
The posterior mean is positive for
destination zones and negative for origin zones. Consistently,
this implies that zones with high levels of accessibility are more
likely to attract trips than low-accessible zones. Conversely,
high-accessible zones are less likely to produce trips than
low-accessible zones. A possible explanation for the negative
origin effect is that high levels of accessibility within a zone
might encourage intra-zonal trips and reduce outgoing trips. Car
ownership is traditionally used as an explanatory variable with
positive impact in trip-production models. In agreement, the
posterior mean for car ownership is positive for origin zones,
which means that zones with high car ownership ratios also have
high trip-production rates. The estimate is negative for
destination zones implying that high car ownership ratios are
negatively correlated with trip-attraction. The negative
destination effect may be attributed to congestion issues.

Distance with parameter $\beta _{25} $ is the final variable.
Distance is a key variable in
gravity-type and direct-demand models, since it is directly
related to the costs of the deterrence function used within the
trip-distribution step. In our model distance has a negative
posterior mean which accords with the basic deterrent
gravitational assumption of trip-distribution models. Furthermore,
based on the posterior mean over standard deviation ratio, distance
is the most significant explanatory variable in all models.


Table \ref{mainresult} also includes the values of the AIC, BIC
and marginal/hierarchical DIC.
The posterior mean of the deviance is used for the calculation of AIC and BIC.
The three criteria provide more support to the PLN
and PIG models,
which provides a
justification for the similarity of the posterior estimates from
the two models.
Furthermore, all three
criteria
indicate that the PIG distribution is the
most appropriate marginal sampling distribution. The hierarchical
DIC is calculated based on
reduced samples of 500 draws, due to memory limitations given the large dimensionality of the data-augmented space.
In addition, sampling the random effects of the PLN model is relatively complicated and time-consuming,
therefore we focus on the PG and PIG models for the remainder of this paper. Based on the hierarchical DIC it is difficult to
distinguish which hierarchical model is more appropriate for predictive purposes, since the differences between
the PG and PIG are marginal.

The random effects present some dissimilarities between the two
models. The range of the PG random effects is from
$\mbox{3.81}\times 10^{ - 8}$ to 40.61 (-17.83 to 3.71 on
log-scale), while the PIG random effects range from $9.48\times
10^{ - 3}$ to 132.35 (-4.66 to 4.89 on log-scale). Due to the GIG
posterior distribution, the PIG random effects exhibit a longer
right-tail than the PG random effects which are gamma distributed.
On logarithmic scale the PIG random effects are relatively more
symmetrical near 0, whereas the PG random effects have a longer
left-tail.

\subsection{Sensitivity analysis for hyperpriors}

\begin{figure}
\begin{center}
 \includegraphics[scale=0.38]{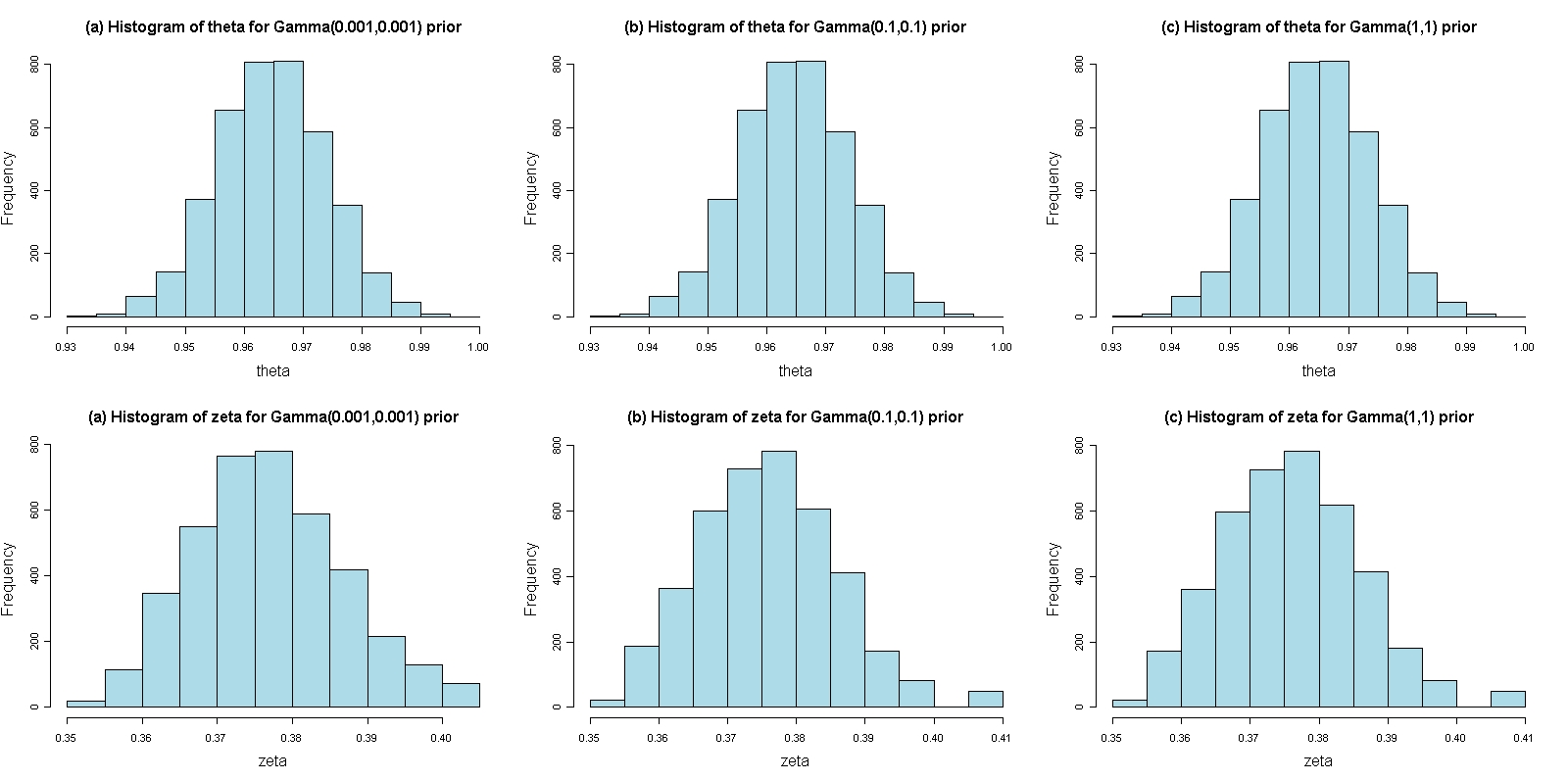}
\caption{\label{dispersion_parameters} Histograms of dispersion
parameters $\theta$ (top) and $\zeta$ (bottom) under gamma
hyperpriors with $a$ equal to; (a) 0.001, (b) 0.1 and (c)
1.}\end{center}
\end{figure}

In this section we perform a sensitivity analysis for parameter
$a$ of the gamma hyperpriors assigned to parameters $\theta$ and
$\zeta$ of the PG and PIG models. Parameter $\sigma^2$ is not
included in the analysis, due to the substantial time which is
required for M-H simulation from the PLN model. Histograms of 4000
posterior draws of parameters $\theta$ and $\zeta$ for values of
$a$ equal to 0.001 (the initial value), 0.1 and 1 are presented in
Figure \ref{dispersion_parameters}.
As seen, the posterior distributions
are not influenced by hyperparameter $a$. One can notice a slight change in the right tail of the posterior distribution of $\zeta$
for $a$ equal to 0.1 and 1, nonetheless this does not affect posterior inferences.
The results are in line with the discussion in
Gelman (2006), since in our case the random effects are observational and therefore we would not expect
to have the sensitivity problems that arise in grouped random effects settings for small numbers of groups.

\subsection{Posterior predictive checks for OD flows}

For overall goodness-of-fit, we employ posterior predictive checks (Meng, 1994) for the absolute and squared
distances
(with respect to the expected values)
and for the
hierarchical deviance.
The absolute distance is more sensitive to small
deviations, whereas squared distance assigns more penalty to large
deviations. Each test quantity is calculated for observed and
predicted data over the 500 posterior draws.

\begin{table}[htbp]
\begin{center}
\begin{tabular}{cccc}
\hline\textbf{Test quantity}& \textbf{Formula}&
\textbf{PG}&
\textbf{PIG} \\
\hline Absolute distance& $\sum {\left( {{\rm {\bf y}} - E({\rm
{\bf y}}\vert {\boldsymbol {\beta }},{\rm {\bf u}})} \right)} $&
0.278&
0.440 \\
 Squared distance& $\sum {\left( {{\rm {\bf y}} - E({\rm
{\bf y}}\vert {\boldsymbol {\beta }},{\rm {\bf u}})} \right)^2} $&
0.532&
0.488 \\
Deviance& $ - 2\log p({\rm {\bf y}}\vert {\boldsymbol {\beta }},{\rm
{\bf u}})$& 0.996&
0.648 \\
\hline
\end{tabular}
\caption{Bayesian p-values for the absolute distance, squared
distance and deviance test quantities from 500 posterior draws of
the hierarchical PG and PIG models.} \label{tab2}
\end{center}
\end{table}

The test quantities
with the corresponding Bayesian p-values are presented in Table
\ref{tab2}. In general, both models provide satisfactory Bayesian
p-values for squared
distances, which are close to the ideal value of 0.5. Predictions from the PIG seem to replicate better the
observed data for small deviations from the expected values and
also with respect to the Poisson distributional assumption. Note that the aim here is not model comparison,
but examination of the characteristics of predictions.

\begin{figure}
\begin{center}
 \includegraphics[scale=0.38]{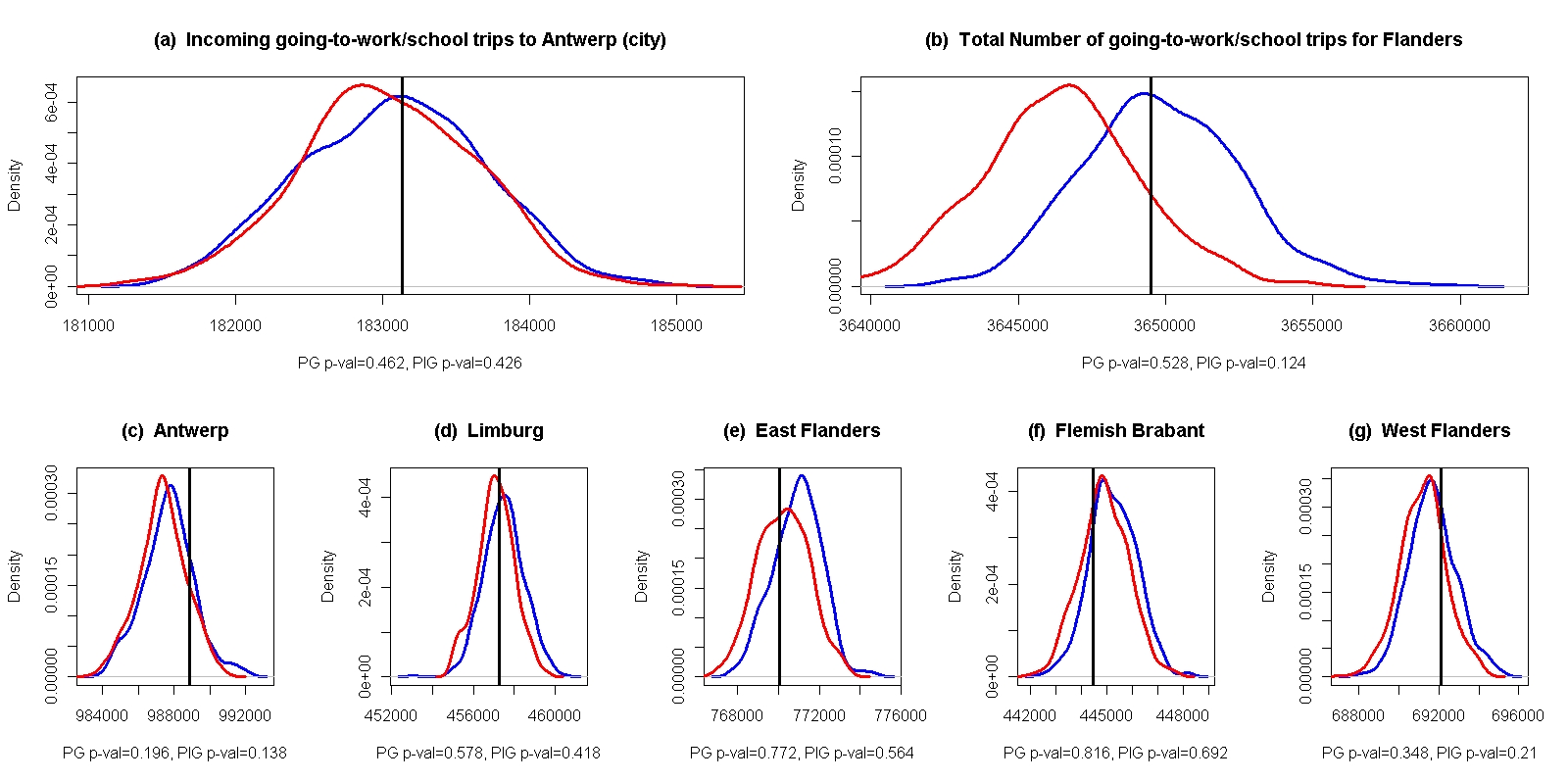}
\caption{\label{kernels} Kernel estimates of predictive
distributions for going-to-work/school trips from the PG model (in
blue) and the PIG model (in red) for (a) incoming trips to the
city of Antwerp, (b) total number of trips in Flanders and
intra-zonal trips for the five Flemish provinces; (c) Antwerp, (d)
Limburg, (e) East Flanders, (f) Flemish Brabant and (g) West
Flanders. The vertical black lines indicate the observed
quantities.}\end{center}
\end{figure}

An interesting feature of OD modeling is that the
administrative structure allows for various aggregations of observed
and replicated data with respect to administrative levels and also
types of trips.  From a
statistical perspective, the aggregated distributions can be
compared to the observed aggregated values, thus resulting in
Bayesian p-values for case-specific tests. Examples of such tests
for incoming trips to the municipality of Antwerp, total trips for
Flanders and intra-zonal trips for the five Flemish provinces
are presented in Figure \ref{kernels}. In general, all p-values
are within acceptable limits. From a transportation planning
perspective, such predictions
are particularly useful for policy-evaluation.

\subsection{Predictive inference for link flows}

For traffic-assignment we utilize the deterministic user
equilibrium (DUE) model which is based on Wardrop's 1$^{st}$
principle (Wardrop, 1952), also known as the \textit{equilibrium
principle}.
In short, DUE
assignment uses an iterative process in order to reach a
convergent solution in which travelers cannot reduce their travel
times by switching routes. At each iteration link capacity
restraints and link flow-dependent travel times are taken into
account in order to calculate link flows. As \textit{link
performance function} we adopt the common BPR formulation (Bureau
of Public Roads, 1964) which relates link travel times to
volume-over-capacity (V/C) ratios, specifically $t = t_f [ {1 +
\alpha ( {v \mathord{/ {\vphantom {v c}}
\kern-\nulldelimiterspace} c} )^\beta }]$, where $t$
is the link travel time, $t_f $ is link free-flow travel time, $v$
is link volume (flow), $c$ is link capacity and $\alpha $, $\beta
$ are calibration parameters which are set equal to their
historical values of 0.15 and 4, respectively. If we denote by
${\rm {\bf A}}$ the DUE assignment operator, we execute 500
individual assignments from the predictive OD's of each model and
obtain 500 corresponding link flow or link volume vectors ${\rm
{\bf v}}$. That is, ${\rm {\bf A}}{\rm {\bf
y}}^{pred(m)} = {\rm {\bf v}}^{(m)}$ for $m = 1,2,...,500$,
where ${\rm {\bf v}}^{(m)} = (v_1^{(m)} ,v_2^{(m)}
,...,v_l^{(m)} )^T$ and $l$ is the total number of network
links. For the Flemish network $l$ is equal to 97450. The
assignments concern the morning peak-hour interval between 7 am
and 8 am for a normal weekday.

\begin{figure}
\begin{center}
 \includegraphics[scale=0.42]{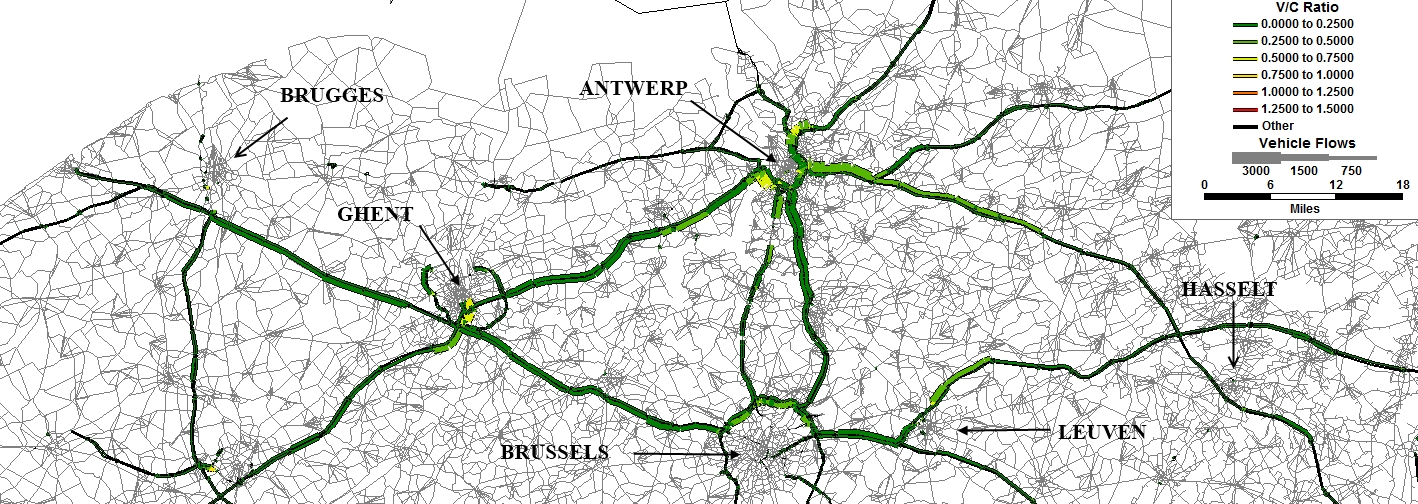}
\caption{\label{flows} Mean visualization of highway flows and V/C
ratios for going-to-work/school trips between 7-8 am in Flanders
under DUE assignment and PIG OD predictions.}\end{center}
\end{figure}

The mean state of the Flemish network under DUE assignment and OD
predictions from the PIG model is presented in Figure \ref{flows}.
By ``mean state'' it is meant that the 500 link flow vectors were
averaged first and the visualized.
In order to make Figure \ref{flows} simpler to comprehend only
volumes and V/C ratios for highway links are highlighted. The main
findings are the following. V/C ratios are higher in specific
segments on or near the highways rings of Antwerp (R1) and Ghent
(R4), which can be identified by the yellow spots indicating V/C
ratios between 0.5 and 0.75. Relatively high V/C ratios (light
green colour) also occur on the northern part of highway ring R0
around Brussels, on highway E40 near Leuven, highway E313 which
connects Antwerp with Hasselt and to a lesser degree on highways
E17 and E19 which connect Antwerp with Ghent and Brussels,
respectively. The corresponding visualization map based on PG
predictions is not presented as it seems almost identical to the
one in Figure \ref{flows}, with differences being difficult to
spot on a global scale.

\begin{figure}
\begin{center}
 \includegraphics[scale=0.38]{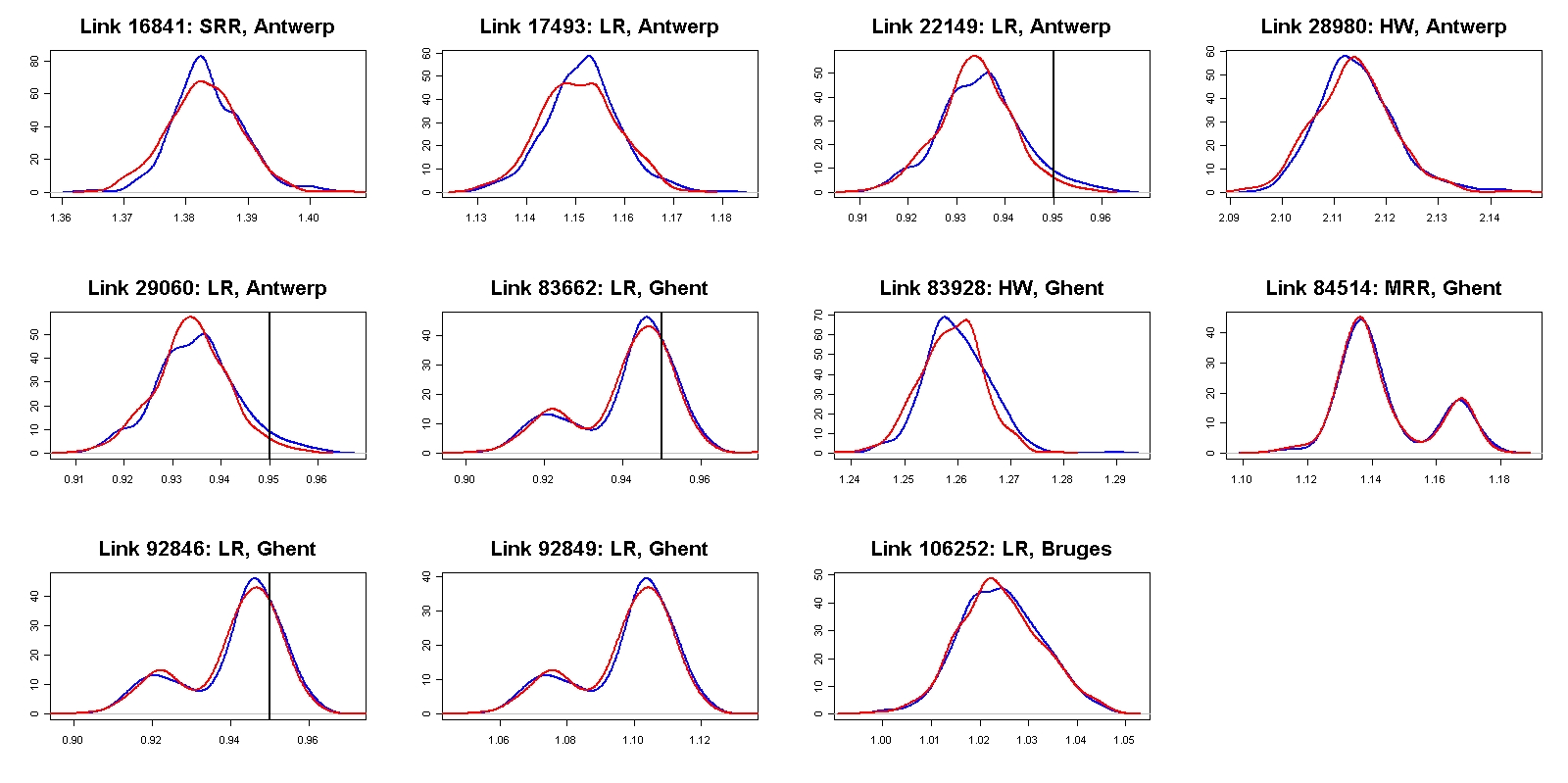}
\caption{\label{kernelPG} Kernel estimates of the PG (in blue) and
PIG (in red) V/C distributions of the 11 congested links which
either include or exceed the threshold value of 0.95 highlighted
by a vertical black line in the distributions which include this
value. The abbreviations HW, MRR, SRR and LR stand for highways,
main regional roads, small regional roads and local
roads.}\end{center}
\end{figure}

An interesting application is the identification of congested
links on the network. Congestion identification is related
to \textit{critical link} identification, which is customarily a
subject of \textit{vulnerability analysis} and relies
significantly on traffic-assignment procedures (see e.g. Jenelius et al., 2006). Through our approach congested links
are evaluated directly in terms of probability estimates. As
congested links we define those links on which the V/C ratio
exceeds a certain threshold value $t$ with a certain probability
$P(\mbox{V/C} > t)$.
As a conservative choice and in order not to overestimate the
number of critical links a threshold value of 0.95 is adopted,
based on the assumption that the majority of trips taking place
between 7 am and 8 am are either work or school related trips. For
$t = 0.95$ congestion is identified in eleven links which all
belong to large Flemish municipalities; 5 in Antwerp, 5 in Ghent
and 1 link in Bruges.
The V/C distributions from both models are presented in Figure
\ref{kernelPG}.

Certain remarks can be made, based on Figure
6, regarding V/C distributions and consequently link flow
distributions from DUE assignment. First, the choice of statistical model does not seem to affect individual
V/C distributions as the corresponding distributions are very similar. Second, individual V/C and link
distributions are not necessarily close to normal distributions --
for instance bimodalities are observed -- in contrast to
aggregated distributions (e.g. link flows for highways -- not presented here) which converge to
normality in accordance to the central limit theorem. Third, the
bimodalities may be attributed to the iterative user equilibrium
procedure; when the flows on a specific link and at a given
iteration exceed a certain threshold -- leading to a high V/C
ratio -- and there exists an alternative link which has a cost
which is close but lower, then in the following iteration a switch
of flows will occur from the high-cost link to the low-cost link.
This ``switching'' effect will eventually result to bimodal
distributions as the ones observed in Figure 6.

\begin{table}[htbp]
\begin{center}
\begin{tabular}{cccccc} \hline
 \textbf{Congested} &  \textbf{Link}& \multicolumn{2}{c}{ \textbf{PG}} &
\multicolumn{2}{c}{ \textbf{PIG}}  \\
  \textbf{link}& \textbf{type}
 &
$E(V / C)$& $P(V / C > 0.95)$& $E(V / C)$&
$P(V / C > 0.95)$ \\
\hline
 16841& Small regional road& 1.384& 1& 1.383&
1 \\
 17493& Local road& 1.152& 1& 1.151&
1 \\
 22149& Local road& 0.935& 0.046& 0.934&
0.022 \\
 28980& Highway& 2.114& 1& 2.114&
1 \\
29060& Local road& 0.935& 0.046& 0.934&
0.022 \\
 83662& Local road& 0.941& 0.236& 0.941&
0.208 \\
 83928& Highway& 1.260& 1& 1.259&
1 \\
 84514& Main regional road& 1.144& 1& 1.144&
1 \\
 92846& Local road& 0.941& 0.236& 0.941&
0.208 \\
 92849& Local road& 1.098& 1& 1.097&
1 \\
106252& Local road& 1.024& 1& 1.024&
1 \\
\hline
\end{tabular}
\caption{\label{tab3}Expected V/C ratios and probabilities of
exceeding a V/C of 0.95 for the 11 congested links under DUE
assignment and PG, PIG predictions.}
\end{center}
\end{table}

Seven out of the eleven links have a V/C value greater than 0.95
with probability 1. Visual examination of the distributions in
Figure 6 additionally reveals that these seven links also exceed
the value $t = 1$ with probability 1, except perhaps of link
106252 which has its minimum located near 1 and may therefore
include smaller values than 1 with a low probability.
The
remaining four links have lower V/C ratios and exceed the value
0.95 with a probability lower than one. The expected values and
the corresponding probabilities for the eleven congested links are
presented in Table \ref{tab3}. Assignment with PG predictions
results to slightly higher probabilities for links 22149, 29060,
83662 and 92846. We also note that if the analysis was based on
the expected values congestion would not have been identified on
those four links.

\section{Discussion}

In this paper we investigated the use of Poisson
mixtures in OD modeling as a viable alternative to traditional
transportation models. The advantages of the proposed approach
are; i) it incorporates the steps of trip-generation and
trip-distribution in statistical models which provide a wider
inferential scope and ii) it allows for probabilistic inference on
link traffic and congestion, conditional on the assignment model.
At the same time, the approach may be viewed as a statistical,
direct-demand, gravity model, thus retaining a strong relation
with traditional transportation models.

The case study focused on a large, sparsely distributed and overdispersed OD matrix derived from the 2001 Belgian travel census covering the region of Flanders. In particular, we considered the PG, PLN and PIG models as alternative modeling options. The PIG -- a model not as popular as its competing alternatives -- provided the best marginal fit and resulted in consistent short-term predictions. Given the convenient distributional properties of the PIG model, we recommend its use when analyzing large-scale OD matrices. In addition, we investigated the performance of INLA compared to MCMC for the PG and PLN models and found that INLA can provide fast and accurate approximations. From this point of view, INLA is particularly suited for the PLN, which proved to be the most cumbersome model to work with using MCMC. Further development of the R-INLA package, in terms of prior extensions, will make it a useful tool for large-scale OD analysis.

Future research directions concerning transportation issues are many.
First, the set of covariates used in this study is by no means conclusive. As pointed out by one referee the models could improve in terms of capturing representation of activities in destination-zones. This can be achieved by including number of workplaces and shopping facilities as predictors. This type of information was not available and could not be included in the current analysis.
Second, the issue of modal-split which was not pursued here can be potentially incorporated in the proposed modeling approach. A third issue concerns dynamic modeling of short-term OD matrices, for example analysis of OD matrices on hourly intervals. A fourth category of issues is related to a series of traffic-assignment comparative studies between the DUE model, utilized here, and other assignment models such as the stochastic user equilibrium model, conditional on Bayesian predictions.

From a statistical perspective, discrete random effects could have
been considered as an alternative approach for clustering purposes.
This approach was not
pursued here
as the focus of this study was on modeling and capturing the heterogeneity per OD pair.
Finally, the PIG model can be of potential value to any other count data analysis problem
under the presence of overdispersion. From this point of view,
it will be interesting to consider zero-inflated model extensions and also
to compare to other mixing/prior distributional designs.

\section*{\\ Acknowledgements}

The authors would like to thank the associate editor and two anonymous referees for their useful comments and suggestions. Moreover, we would like to thank the INLA support team, particularly, Prof. H\aa vard Rue for his suggestions and technical help concerning implementation of INLA.

\newpage

\section*{\\ Appendix A: Metropolis-Hastings simulation}

We utilize M-H simulation on the marginal structures
in order to bypass sampling 94864
random effects at each MCMC iteration.
Although sampling
\textbf{u} in a Gibbs-like fashion is straightforward for the hierarchical PG
and PIG models, memory limitations
would require discarding \textbf{u} at the end of each iteration.
M-H for the marginal PG and PIG structures
is far more efficient with ${\boldsymbol {\beta }}$, $\theta $ and
$\zeta $ being easy to sample, while \textbf{u} can be generated
subsequently as described in sections 2.1 and 2.3. The PLN model
is more problematic since an additional Metropolis step or
rejection sampling is required for the hierarchical
structure, which is an obvious burden for 94864 random effects. On
the other hand, simulation for the marginal PLN structure requires
numerical or MC integration within MCMC and -- in addition --
vector \textbf{u} is not easy to sample subsequently.

In particular, we employ an independence-chain M-H algorithm where
the location and scale of the proposals are fixed (see e.g. Chib and Greenberg, 1995) to the
corresponding ML estimates. For regression vector ${\boldsymbol {\bf \beta
}}$ a multivariate normal proposal is used, i.e. $q({\boldsymbol {\bf
\beta }}) = {\rm {\bf N}}_{p + 1} ({\boldsymbol {\beta }}^{ML},{\rm
{\bf V}}_{\boldsymbol {\beta }}^{ML} )$ with ${\boldsymbol {\beta }}^{ML}$
being the ML estimate of ${\boldsymbol {\beta }}$ and ${\rm {\bf
V}}_{\boldsymbol {\beta }}^{ML} $ the estimated variance-covariance
matrix of ${\boldsymbol {\beta }}^{ML}$ for each model. For the
dispersion parameters $\theta $, $\sigma^2 $ and $\zeta $ we
used the following gamma proposals; $q(\theta ) = Gamma(a_{PG}
,b_{PG} )$, $q(\sigma^2 ) = Gamma(a_{PLN} ,b_{PLN} )$ and
$q(\zeta ) = Gamma(a_{PIG} ,b_{PIG} )$ with proposal parameters
set to satisfy the conditions $a_{PG} / b_{PG} = \theta^{ML}$,
$a_{PG} / b_{PG}^2 = Var(\theta^{ML})$, $a_{PLN} / b_{PLN} =
\sigma^{2^{ML}}$, $a_{PLN} / b_{PLN}^2 = Var(\sigma^{2
^{ML}})$, $a_{PIG} / b_{PIG} = \zeta ^{ML}$ and $a_{PIG} /
b_{PIG}^2 = Var(\zeta ^{ML})$.
Regarding probability
calculations from the PLN distribution we implemented both
numerical and MC integration. Results showed that the MC sample $L$
should be preferably 2000 in order to obtain stable estimates, similar to the estimates from numerical integration, while
numerical integration was already two-times faster than MC
integration with a sample of 200. Therefore, numerical integration
was preferred.

We utilized 5 independent M-H chains of size 4200 and discarded the first 200 iterations as burn-in,
resulting in posterior samples of 20000 draws.
The 10$^{th}$, 30$^{th}$, 50$^{th}$, 70$^{th}$ and 90$^{th}$ percentile points of
the proposal distributions were used starting values.
The resulting
acceptance ratios were 72{\%} for the PG model, 67{\%} for the PLN
model and 33{\%} for the PIG model, on average.
The multi-chain diagnostics of Gelman and Rubin (1992) and Brooks and Gelman (1998) were used to asses convergence.
All univariate potential scale reduction factors (PSRF) were very close to 1 for all 3 models.
The multivariate PSRF for the PG, PLN and PIG were 1.01, 1.01 and 1.06, respectively.
Finally, in order to reduce the computational burden of the subsequent analysis
the posterior samples were thinned by an interval of 5, resulting in final posterior samples
of 4000 draws.

Implementation of MCMC was done in R version 2.8.2 on a 64bit Windows Server 2003 R2 with 32 GB of RAM.
The simulations for the PG and PIG models
required approximately 1 and 2.4 hours, respectively, whereas the
PLN model required 3.6 days due to numerical integration.

The INLA models based on the prior assumption $ {\boldsymbol {\beta }}\sim {\rm {\bf N}}_{26} ({\rm {\bf 0}},{\rm {{{\bf\ I}_{26} \sigma^2 }}}), \mbox{ with } \sigma^2=10^3$, were fitted remotely in R version 3.0.1 through the Linux Server
maintained by the INLA support team. The PG model required approximately 2.2 hours and the PLN model about 2.5 hours.
The Gaussian approximation was used for both models.

\section*{\\ Appendix B: INLA estimates}
Here we present the INLA estimates for the PG and PLN models using
the Gaussian approximation for  the entire dataset.

\begin{center}
\begin{tabular}{ccccc}
\hline
\multirow{2}{*}{\textbf{Parameter}} & \multicolumn{2}{c}{\textbf{PG}} & \multicolumn{2}{c}{\textbf{PLN}}\tabularnewline
 & Mean & 95\% Cr. Int. & Mean & 95\% Cr. Int.\tabularnewline
\hline
$\beta_{0}$ & 3.605  & (2.903, 4.307) & 5.974 & (5.106, 6.847)\tabularnewline
$\beta_{1}$ DP & 0.005  & (0.005, 0.005) & 0.005  & (0.005, 0.005)\tabularnewline
$\beta_{2}$ DA & 0.007  & (0.006, 0.007) & 0.007  & (0.006, 0.007)\tabularnewline
$\beta_{3}$ DD & 0.008  & (0.007, 0.008) & 0.008  & (0.008, 0.009)\tabularnewline
$\beta_{4}$ DC & 0.008  & (0.007, 0.008) & 0.007  & (0.006, 0.008)\tabularnewline
$\beta_{5}$ DM & -0.082  & (-0.084, -0.081) & -0.078  & (-0.080, -0.076)\tabularnewline
$\beta_{6}$ DE & 0.435  & (0.404, 0.466) & 0.509  & (0.471, 0.547)\tabularnewline
$\beta_{7}$ MC & 0.459  & (0.427, 0.491) & 0.431  & (0.392, 0.470)\tabularnewline
$\beta_{8}$ MD & -0.482  & (-0.523, -0.442) & -0.424  & (-0.472, -0.376)\tabularnewline
$\beta_{9}$ MA & -0.079  & (-0.109, -0.048) & -0.177  & (-0.214, -0.139)\tabularnewline
$\beta_{10}$ MP & -0.447  & (-0.569, -0.324) & -0.699  & (-0.848, -0.551)\tabularnewline
$\beta_{11}$ ER(o) & -1.037  & (-1.163, -0.911) & -0.458  & (-0.603, -0.313)\tabularnewline
$\beta_{12}$ ER(d) & 0.310  & (0.192, 0.428) & 0.449  & (0.302, 0.596)\tabularnewline
$\beta_{13}$ PD(o) & 0.495  & (0.471, 0.519) & 0.469 & (0.440, 0.498)\tabularnewline
$\beta_{14}$ PD(d) & 0.558  & (0.533, 0.583) & 0.591  & (0.560, 0.622)\tabularnewline
$\beta_{15}$ RL(o) & -0.315  & (-0.353, -0.278) & -0.299  & (-0.344, -0.254)\tabularnewline
$\beta_{16}$ RL(d) & 0.287  & (0.250, 0.324) & 0.253  & (0.208, 0.298)\tabularnewline
$\beta_{17}$ PL(o) & 1.272  & (1.232, 1.312) & 1.212  & (1.165, 1.259)\tabularnewline
$\beta_{18}$ PL(d) & 0.442  & (0.402, 0.482) & 0.479  & (0.429, 0.528)\tabularnewline
$\beta_{19}$ CR(o) & 3.401  & (3.140, 3.663) & 3.183  & (2.871, 3.495)\tabularnewline
$\beta_{20}$ CR(d) & -1.549  & (-1.813, -1.286) & -1.177  & (-1.496, -0.859)\tabularnewline
$\beta_{21}$ HT(o) & 0.010  & (0.007, 0.012) & 0.010  & (0.007, 0.014)\tabularnewline
$\beta_{22}$ HT(d) & 0.051  & (0.048, 0.054) & 0.047  & (0.043, 0.050)\tabularnewline
$\beta_{23}$ PMT(o) & 0.264  & (0.246, 0.281) & 0.263  & (0.242, 0.283)\tabularnewline
$\beta_{24}$ PMT(d) & 0.867  & (0.850, 0.884) & 0.824  & (0.802, 0.846)\tabularnewline
$\beta_{25}$ D & -2.912  & (-2.930, -2.893) & -2.807  & (-2.830, -2.785)\tabularnewline
$\theta$ & 0.969 & (0.950, 0.986) & \multicolumn{2}{c}{-}\tabularnewline
$\sigma^{2}$ & \multicolumn{2}{c}{-} & 1.020 & (1.000, 1.050)\tabularnewline
\hline
\end{tabular}
\par\end{center}

\newpage

\bibliographystyle{plainnat}

\end{document}